\documentclass[11pt,a4paper,twoside]{article}      
\usepackage{amsmath,amssymb,amsfonts} 
\usepackage{graphics}                 
\usepackage{color}                    
\usepackage{hyperref}                 
\usepackage{latexsym}
\usepackage{verbatim}
\usepackage{graphicx}

\begin{document}
\date{\mbox{ }}
\title{{\normalsize  IPPP/10/59DCPT/10/118\hfill\mbox{}\hfill\mbox{}}\\
\vspace{2.5 cm}
\LARGE{\textbf{Gauge Mediation with Gauge Messengers in SU(5)}}}
\author{Luis Matos\\[2ex]
\small{\em Institute for Particle Physics Phenomenology,}\\
\small{\em Durham University, Durham DH1 3LE, United Kingdom}\\[2ex]}
\date{}
\maketitle

\vspace{10ex}

\begin{abstract}
\noindent
The inclusion of gauge messengers in models of gauge mediation allows for more general predictions that those described by the framework of general gauge mediation. Motivated by this, we explore some models of gauge mediation with gauge messengers in SU(5) GUTs.

In most previous attempts of building viable models where gauge messengers play a role in determining the soft terms, squark and/or slepton masses turned out to be tachyonic. The objective of this paper is to address this problem and propose two possible solutions, one of which has a natural realization in the solution of the doublet-triplet problem.

Another interesting result is that in these models the association of SUSY breaking with the breaking of the GUT group provides a simple mechanism that can explain why $SU(5)\rightarrow  SU(3)\times SU(2) \times U(1)$ is preferred over other symmetry breaking patterns.

\end{abstract}

\newpage

\section{Introduction:}

A lot of work has been done in in recent years trying to understand the most general way to describe the sort of mass spectra one can expect to find at the LHC if SUSY is realized in nature and is communicated to the MSSM through gauge interactions (\cite{Intriligator:2007py,Meade:2008wd,Buican:2009vv,Abel:2009ve,Abel:2007nr,Abel:2008gv,Carpenter:2008wi,Dine:2006xt,Dine:2007dz,Intriligator:2010be,Dermisek:2006qj,Giudice:1997ni,McGarrie:2010kh,Seiberg:2008qj,Benakli:2010gi}). A framework known as general gauge mediation \cite{Meade:2008wd} (GGM) was constructed and, under very general assumptions, it describes (in a model independent way) the possible set of soft terms one can expect with only a small number of parameters. Sum rules for the squark/slepton masses were also derived.

From this work other features that were necessary for a particular model to be viable have emerged. One example is the requirement that the universe should be in a perturbative metastable vacuum \cite{Nelson:1993nf,Shih:2007av,Komargodski:2009jf}. This conclusion arose from the fact that if this is not assumed, gaugino masses vanish at leading order in $\frac{F}{M}$. This would generally give a small hierarchy between gaugino masses and the soft scalar masses, which leads to some tuning to get the correct electroweak symmetry breaking scale.\footnote{A more general argument that any vacuum of a model with low energy SUSY breaking model should be metastable can be made by saying that any spontaneously broken R-symmetry should be only approximate in order not to have a massless R-axion. This is a much weaker assumption: it only requires that SUSY is restaured somewhere in field space (e.g. non-perturbativelly), while the previous argument requires that this vacuum must be accessible within the regime of validity of perturbation theory.}

In \cite{Buican:2009vv}, it was noted that the predictions of GGM could be generalized by allowing the presence of gauge messengers. These are Higgsed vector fields that couple to SUSY breaking vevs and thus act as messengers. An extension of GGM which allowed for gauge messengers was constructed in \cite{Intriligator:2010be}. One particular difference is that gaugino masses are generated at leading order in $\frac{F}{M}$ even if the vacuum is not metastable.

This scenario is then richer than the one described by GGM: it allows for a more general class of soft terms and the sum rules of GGM are changed \cite{Buican:2009vv}. The main motivation of this work is then to explore the role of gauge messengers and see whether these models lead to qualitatively different conclusions.

One problem that arose when trying to build models of this type is that some of them have tachyonic squark and/or slepton masses \cite{Giudice:1997ni,Intriligator:2010be,Buican:2009vv}. This is because the leading order (one loop) contribution to the soft scalar masses is always tachyonic. In \cite{Intriligator:2010be} was shown that this contribution can be suppressed with respect to the two loop corrections. But even these are often negative, so that squarks/sleptons remain tachyonic even at two loops.

The aim of this paper is then to address this problem, and propose possible solutions.

Its structure is as follows: in section 2 we give a description of the basic model we will be considering. The messengers will be adjoints of $SU(5)$. We will show that in the vacuum, and because we will couple these adjoints to some F-terms, the $SU(5)$ is naturally broken to the MSSM gauge group $SU(3)\times SU(2)\times U(1)$. This is different from many models where the GUT breaking respects SUSY and the choice of symmetry breaking pattern is chosen.

In section 3 is a review of the results of \cite{Intriligator:2010be,Buican:2009vv}.

In section 4 we explore the problem of tachyonic squarks and sleptons and show that in a large class of models the sign of the soft masses depends only on the field content of the messenger sector, and not on the parameters: $F/M$ turns out to be universal for both gauge and non-gauge messengers.

In section 5 we show how the constraints used to derive the previous result can be evaded with two examples, and we present the conclusions in section 6.

\section{A toy model:}

The model for the messenger sector consists of two fields that are adjoints under the $SU(5)$ GUT and a singlet. The superpotential is given by:
\begin{equation}
 W=-\mu^2 \Phi+\lambda \Phi Tr(Y_0 Y_0)+mTr(Y_2 Y_0) +\overline{\lambda}Tr(Y_2 Y_0 Y_0)
\end{equation}
Where $\Phi$ is the singlet and $Y_0$ and $Y_2$ are adjoints of $SU(5)$. This superpotential has an R-symmetry such that $R(Y_0)=0$ and $R(Y_2)=R(\Phi)=2$, and is general. The Y-fields can be written using the generators of the Lie Algebra of $SU(5)$: $Y_{i}=Y_{i}^{(a)}T^{(a)}$, where $T^{(a)}$ are the generators (mode details in the appendix A). Note that linear terms in the Y-fields vanish as the generators of $SU(5)$ are traceless.

Let us start by analyzing this model in the limit where $\lambda=0$:
\begin{equation}
 W=-\mu^2 \Phi+mTr(Y_2 Y_0) +\overline{\lambda}Tr(Y_2 Y_0 Y_0)
\end{equation}
 In this case there are two independent sectors: one composed by the singlet and it's superpotential (singlet sector), and the second consisting of the adjoint and its superpotential (adjoint sector). SUSY is broken in the singlet sector as $F_{\Phi}=\mu^2$, and $\Phi$ is the Goldstino. The adjoint sector can have several SUSY solutions:

\begin{itemize}
\item $Y_2=0$, $Y_1=0$;\\
\item $Y_2=0$, $Y_0=\frac{m}{3\overline{\lambda}} \ diag(\{1,1,1,1,-4\})$;\\
\item $Y_2=0$, $Y_0=\frac{m}{\overline{\lambda}}\ diag(\{2,2,2,-3,-3\})$;\\
\end{itemize}

 Where we use the notation $diag(\{x_1,x_2,...,x_n\})$ to denote a diagonal matrix with elements $x_1,...,x_n$. Since SUSY is not broken in the adjoint sector the degeneracy between these vacua is not lifted and (ignoring SUGRA corrections) all should be considered on equal footing.

Let us consider now the beta function associated with the GUT gauge group: the extra adjoints give a very large negative contribution above their mass threshold:
\begin{align}
 b'=&b_{MSSM}-S_{messengers}=3\times N_c-S_{matter}-S_{messengers}=\nonumber\\
&3\times5-3\times \frac{3}{2}-3\times \frac{1}{2}-1-2\times5=-2
\end{align}
Where we take as matter content: the MSSM \cite{Raby:2006sk,Raby:2008gh} (a fundamental $\overline{5}$, an anti-symmetric $10$ and two Higgs, $5$ and $\overline{5}$) and two adjoints for the messenger sector.

Above the GUT scale the gauge coupling will not be asymptotically free. This means that we are implicitly assuming that above the GUT scale the MSSM is actually the dual low energy description os some theory valid at energies well above the GUT scale (other examples where the MSSM is considered to be the dual of another theory are considered in \cite{Abel:2009bj,Abel:2008tx}). In this context, the singlets of the low energy theory could be thought of as composites of some other fields.

Now let us turn on the $\lambda$ parameter:
If $\lambda$ is small enough the solutions will change by a small amount and in the minimum the symmetry breaking pattern should be one of the exhibited by the previous solutions (a more detailed discussion of the general minimization of the potential is presented in the appendix A).

At leading order in $\lambda$ one gets:
{\renewcommand{\arraystretch}{1.8}
\renewcommand{\tabcolsep}{0.4cm}
\begin{table}[h]
\begin{center}
\begin{tabular}{|c | c |}
\hline
$V$        & $Solutions$\\
\hline
\hline

$\mu^4$ & $\begin{array}{c}
       Y_2=0\\
       Y_0=0\\
       \Phi=y\\
           \end{array}$\\
\hline

$\mu^4-\frac{40}{9} \lambda \frac{m^2 \mu^2}{\overline{\lambda}^2}$ & $\begin{array}{c}
       Y_2=y\ diag({1,1,1,1,-4})\\
       Y_0=\frac{m}{3\overline{\lambda}}(1+2\lambda(\frac{\mu}{m})^2) diag(\{1,1,1,1,-4\})\\
       \Phi=y\frac{3\overline{\lambda}}{2\lambda}+3y\frac{\overline{\lambda}\mu^2}{m^2}-\lambda y(\frac{20}{3\overline{\lambda}}+\frac{18 \overline{\lambda} \mu^4}{m^4})\\
           \end{array}$\\
\hline

$\mu^4-60 \lambda \frac{m^2 \mu^2}{\overline{\lambda}^2}$ & $\begin{array}{c}
       Y_2=y\ diag(\{1,1,1,-3/2,-3/2\})\\
        Y_0=\frac{2m}{\overline{\lambda}}(1+2 \lambda (\frac{\mu}{m})^2)diag(\{1,1,1,-3/2,-3/2\})\\
       \Phi=y\frac{\overline{\lambda}}{4\lambda}+y\frac{\overline{\lambda}\mu^2}{2m^2}-y\lambda (\frac{15}{\overline{\lambda}}+\frac{3 \overline{\lambda} \mu^4}{m^4})\\
           \end{array}$\\
\hline
\end{tabular}
\end{center}
\end{table}}

Where, as expected, there is a flat direction associated with the fields with R-charge 2, parameterized here by y. An important point to make is that there are no new complex phases associated with these vevs, so there are no new sources of CP violation.

We noted that when $\lambda=0$ there were three possible solutions and that they all should be considered on equal footing. By coupling the adjoint sector to the singlet sector (SUSY breaking sector), this degeneracy is broken. By choosing $\lambda$ to be positive we see that the preferred vacuum is the one required in the MSSM and none of the others.

The non-vanishing F-terms are:
\begin{eqnarray}
F_{\Phi}=\frac{\partial W}{\partial \Phi}=-\mu^2+30 m^2 \frac{\lambda}{\overline{\lambda}^2}\nonumber\\
F_{Y_2^{(23)}}=\frac{\partial W}{\partial Y_2^{(23)}}=-5\sqrt{3} \mu^2 \frac{\lambda}{\overline{\lambda}^2}\\
F_{Y_2^{(24)}}=\frac{\partial W}{\partial Y_2^{(24)}}=-3\sqrt{5} \mu^2 \frac{\lambda}{\overline{\lambda}^2}\nonumber
\end{eqnarray}
Where:
\begin{align}
&T^{23}=diag({1/(2\sqrt{6}),1/(2\sqrt{6}),1/(2\sqrt{6}),-3/(2\sqrt{6}),0})\nonumber\\ &T^{24}=diag({1/(2\sqrt{10}),1/(2\sqrt{10}),1/(2\sqrt{10}),1/(2\sqrt{10}),-2/\sqrt{10}}) \nonumber
\end{align}
So, some of the F-terms are not invariant under the GUT gauge group. As we will see this is a necessary and sufficient condition for the existence of gauge messengers.

\subsubsection{Gauge messengers:}

Now let us consider the gauge messengers. The form of a fermionic mass matrix squared for a generic superpotential and field content is given by:
\begin{equation}
 m^2_f=\left(\begin{array}{cc}
  W_{ij}.W^{jk}+2D^a_iD^{a,k}   &  \sqrt{2}D^{a,j}_iW_{j} \\
 \sqrt{2}D^{b,k}_{,l}W^{l} & D^b_lD^{a,l}+D^a_lD^{b,l}\\
 \end{array}
\right)
\end{equation}
Where $W_{i} (W_{ij})$ is the first (second) derivative of the superpotential with respect to the fields $\phi_i$ (and $\phi_j$), $D^{a}$ is the D-term: $D^a=g \sum_i \phi_i^{\dagger} T^{a} \phi_i$ and $D^{a}_{,j}$ is its derivative with respect to $\phi_j$. Indices are raised and lowered by complex conjugation.\footnote{Gauge invariance of the superpotential in the form of: $D^{a,j}_{,k}W_{j}+D^{a,j}W_{jk} \equiv 0$ was used to simplify the mass matrix.}

In the usual models of gauge mediation the only non-vanishing F-term is associated with a gauge invariant direction. This means that the off-diagonal terms $\sqrt{2}D^{a,j}_iW_{j}$ in the fermionic mass matrix (squared) vanish, so that the only fields that feel the effects of SUSY breaking are the scalars.

This doesn't have to happen: in our case, when $\lambda$ is not zero, there are some F-terms that are not gauge invariant under the full GUT group. This means that the Higgsing of the vector superfields is not SUSY: there will be some mixing between the Higgsed gauginos and the fermionic components of messengers. The mass spectrum of the components of these Higgsed vector superfields will not be SUSY, and because of this they will act as (gauge-)messengers.

For this particular case, it's the bifundamentals of the $SU(3)\times  SU(2)$ that are Higgsed and act as gauge messengers\footnote{See Table \ref{fielddefinitions} for an detailed explanation of the notation used.}.

The mass matrix for these fields at leading order in $\lambda$ is:
\begin{equation}
\left(\chi^{\dagger}, \tau^{\dagger}, \psi^{\dagger}\right) (M_f^{bif.})^2 \left(\begin{array}{c}
                                                         \chi\\
\tau\\
\psi
                                                        \end{array}
\right)
\end{equation}

\begin{equation}
 (M_f^{bif.})^2=\left(\begin{array}{ccc}
\frac{25(m^2+4\lambda \mu^2)g^2}{\overline{\lambda}^2} & \frac{25(m^2+2\lambda \mu^2) g^2}{2 m \overline{\lambda}} y & 0\\
 \frac{25(m^2+2\lambda \mu^2) g^2}{2 m \overline{\lambda}} y  & \frac{25 g^2}{4} y^2  & \frac{10i \lambda \mu^2 g}{\overline{\lambda}} \\
0  & - \frac{10i \lambda \mu^2 g}{\overline{\lambda}}  & \frac{25(4 m^2+16\lambda \mu^2 + \overline{\lambda}^2 y^2) g^2}{4 \overline{\lambda}} y^2  \\
\end{array}
\right)
\end{equation}
So we can see that the F-terms couple to the gauge fields at tree-level and the model has gauge messengers.
The fermionic masses are approximately given by:
\begin{align}
 m_{g.m.,\pm}^2&=25 \frac{g^2}{\overline{\lambda}^2}m^2+\frac{25}{4} y^2 g^2+100 \lambda \frac{g^2}{\overline{\lambda}^2}\mu^2\pm10 g\lambda \frac{y}{\sqrt{4m^2+(\overline{\lambda}y)^2}}\mu^2 \\
m_l^2&=\frac{(8m^2\overline{\lambda}y+(\overline{\lambda}y)^3)^2\lambda^2\mu^4}{m^4(4m^2+(\overline{\lambda}y)^2)}
\end{align}
And $m_{g.m., \pm}$ are the masses of the gauge messengers\footnote{To compute the eigenvalue $m_l$ we computed the fermionic mass matrix to order $\lambda^3$ and then extracted the eigenvalues, which we then expanded to order $\lambda^2$.}.

We note that even though this is F-term breaking the mass splittings come from (tree-level) Kahler potential interactions, not from the superpotential. This is the main difference from the usual models of gauge mediation.

\section{The soft terms:}

The soft terms for a model of gauge mediation with gauge messengers has been recently computed in \cite{Intriligator:2010be,Buican:2009vv,Dermisek:2006qj}. We shall present a short review of these calcultions\cite{Intriligator:2010be} using only wave-function renormalization techniques \cite{Giudice:1997ni,ArkaniHamed:1998kj}.

The main differences from the usual scenarios of gauge mediation are:
\begin{itemize}
 \item gaugino masses are generated at leading order in $\frac{F}{M}$ expansion (even without metastability);
\item soft scalar masses are generated at one loop;
\item trilinear couplings are generated at one loop even at the messenger scale;
\end{itemize}

\subsection{Gaugino Masses:}

\begin{figure}[h!]
\begin{center}
 \begin{tabular}{cccc}
$m_{\lambda}\supset$ &
\begin{tabular}[c]{c}
\includegraphics[scale=0.4]{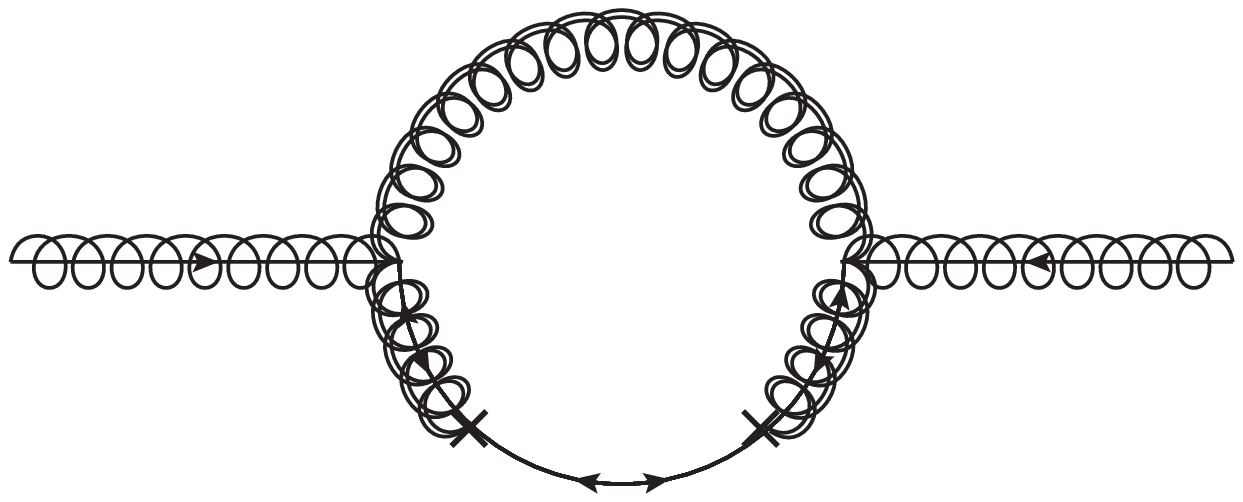}
\end{tabular}
& $+$
 &\begin{tabular}[c]{c} \includegraphics[scale=0.4]{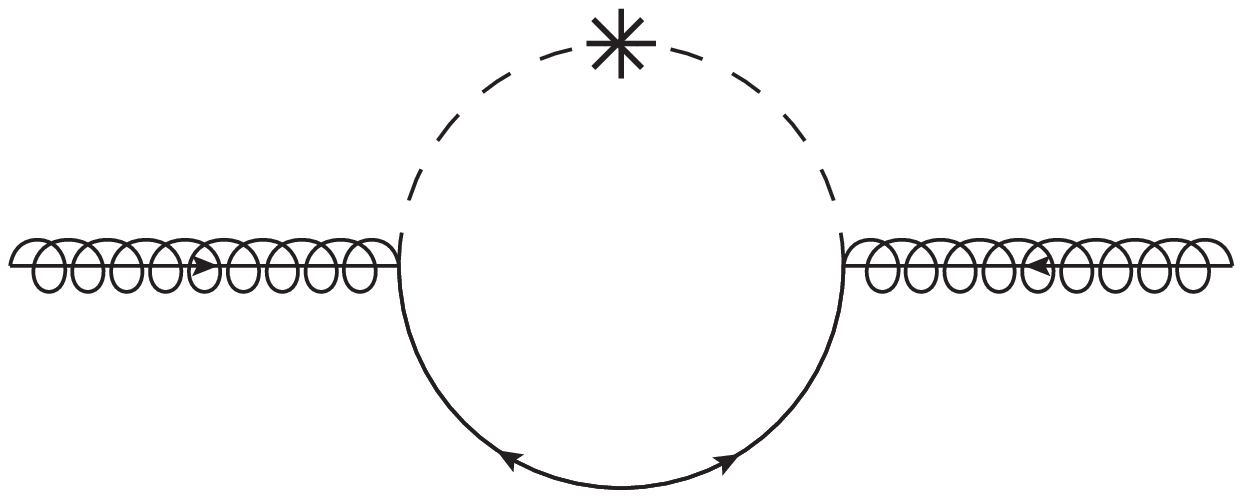}
\end{tabular}
\end{tabular}
\caption{Where the external legs correspond to the un-Higgsed MSSM gauginos, and the doubled wavy-solid (wavy) lines are Higgsed gauginos (gauge bosons), solid (dashed) lines are messenger fermions (scalars). A cross is a mass insertion and a double cross is an F-term insertion. These diagrams give the leading order in F/M gaugino mass contribution.}
\label{gauginodiagrams}
\end{center}
\end{figure}

Diagrammatically we have that the contributions to gaugino masses are given in Figure \ref{gauginodiagrams}. In most models of gauge mediation the second diagram actually doesn't give any contribution.

We will now calculate this contribution: Let us assume that we've fixed some useful gauge (eg. unitary gauge) to perform the calculations and that we call our goldstino field X, so that in the vacuum $<X>=x+\theta^2 F$. Unlike in the usual scenario we will allow x to not be a gauge singlet. So F dictates the scale of SUSY breaking while x is one of the vevs responsible for the breaking of the GUT gauge group to the MSSM gauge groups.

For now let us set F to 0 and work in the SUSY limit.

For the unbroken gauge multiplet, at a given scale $\mu$, the Lagrangian interaction is determined by the X-dependent gauge function $S(X,\mu)$, (where the reason why S can only depend on X is that it must be a holomorphic function of X):
\begin{equation}
 L\supset \int d^2 \theta^2 S(x,\mu) W^{a \alpha}W^a_{\alpha}+h.c.
\end{equation}
If we now turn on F a little bit ($F/M^2<<1$), and since the dependence of S on $<X>$ is holomorphic, the only way that S can change (at one loop) is by the change $<X> \rightarrow X$, where X is now a spurion field. This replacement is called analytical continuation into superspace, since the continuation of $<X>$ to superspace induces a continuation of both gauge coupling and wave-function renormalization to superspace as well (from their dependence on X). The validity of the procedure relies on the fact that this continuation gives the correct R.G. equations for the soft terms.

And S is given by:
\begin{equation}
 S(x,\mu)=\frac{\alpha^{-1}(x,\mu)}{16\pi}-\frac{i\Theta}{32\pi^2}
\end{equation}
Where $\Theta$ is the topological vacuum angle.

So, to compute gaugino masses at leading order in F we need only to solve the R.G. equations for the gauge coupling in the SUSY limit and the continue them to superspace. We note that even though S is holomorphic in the goldstino field, $\alpha^{-1}=16\pi Re(S)$ is not. The one loop R.G. equation is
\begin{equation}
 \frac{d}{dt}\alpha^{-1}=\frac{b}{2\pi}
\end{equation}
where $t=Log(\mu)$ and $b=3N_c-N_f$ for an $SU(N)$ theory. Let us call $b'$ the $\beta$-function coefficient in the U.V. (i.e. above the GUT scale), and $b_i$ the $\beta$-function coefficient of the i-th gauge group below the GUT scale (so $SU(2)$, $SU(3)$ or $U(1)$).
So that the expressions for the holomorphic gauge coupling below the messenger scale is given by:
\begin{equation}
S(\mu)=S(\Lambda_{U.V.})+\frac{b'}{32\pi}Log(\frac{\mu}{\Lambda_{U.V.}})
S_a(\mu)=S(\Lambda_{U.V.})+\frac{b'}{32\pi^2}Log(\frac{X}{\Lambda_{U.V.}})+\frac{b_a}{32\pi^2}Log(\frac{\mu}{X})
\end{equation}
Where $\Lambda_{U.V.}$ is some U.V. cutoff.

The gaugino mass (at this order)\footnote{the gaugino mass, being an observable, depends on the physical gauge coupling, not the holomorphic one. However at one loop there is no difference}is given by
\begin{equation}
 m_{\lambda}=g^2(\mu)S|_{\theta^2}
\end{equation}
So, below the scale $x=<X>$ the gaugino mass is given by the $\theta^2$ component of the gauge function S which is:
\begin{equation}
 m_{\lambda_a}=\frac{\alpha_a(\mu)}{4\pi}(b-b')\frac{F}{x}
\end{equation}
This generalizes for multiple mass thresholds (as long as $F/M^2<<1$). However, it is well known that if the hidden sector superpotential is a cubic polynomial in the fields and one is sitting at the global minimum, the contribution from normal messengers vanishes to leading order in F/M. which means that the only possible non-vanishing contribution is from gauge messengers, and this is given by:
\begin{equation}
 m_{\lambda_a}=-\frac{\alpha_a(\mu)}{2\pi}(N_c'-N_{c,a})\frac{F}{x}
\end{equation}
Where $N_c',N_{c,a}$ are the number of colors in the GUT, MSSM ``a-th'' gauge group, and $-2(N_c'-N_{c,a})$ is the contribution from the Higgsed vector superfields ($(N_c'-N_{c,a})$ from the eaten would be Goldstone Bosons and -3$(N_c'-N_{c,a})$ from the vector field).
Explicit computations for the toy model at hand have been done and and it has been checked that this contribution is non-zero.

\subsection{Scalar Masses:}

\begin{figure}[!h]
\begin{center}
 \begin{tabular}{cccc}
 $m^{(1)}_{Q}\supset$ &
\begin{tabular}[c]{c}
\includegraphics[scale=0.4]{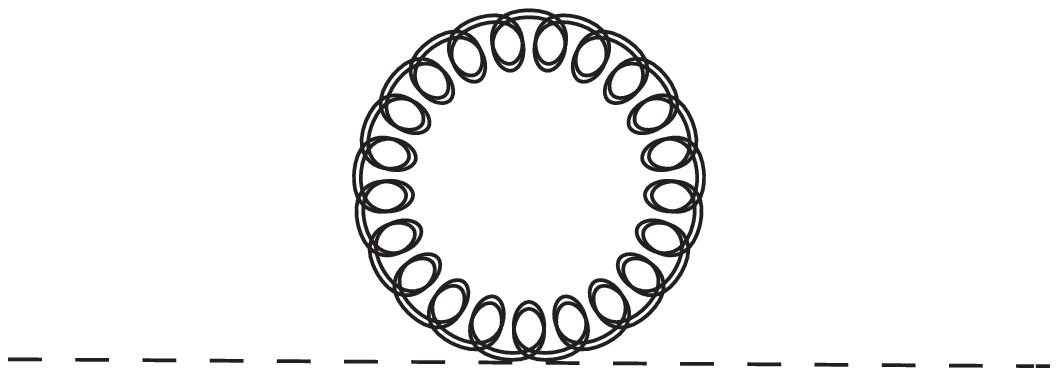}
\end{tabular}
& $+$
 &\begin{tabular}[c]{c}
\includegraphics[scale=0.4]{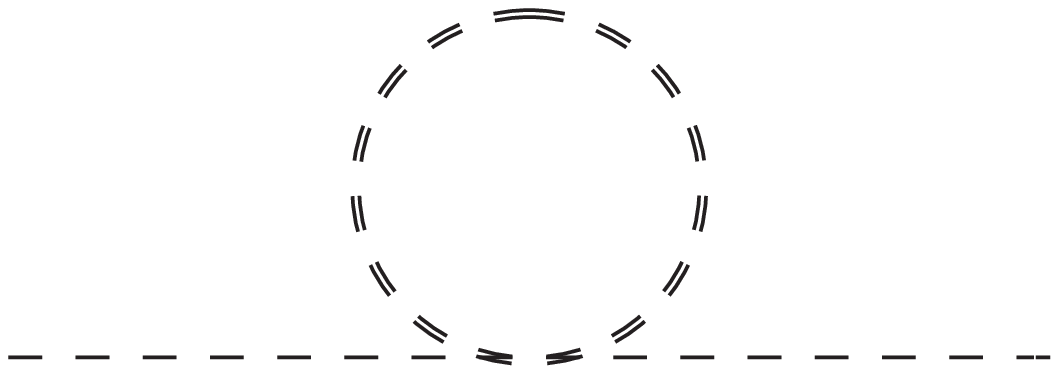}
\end{tabular}\\
  &  \begin{tabular}[c]{c}
\includegraphics[scale=0.4]{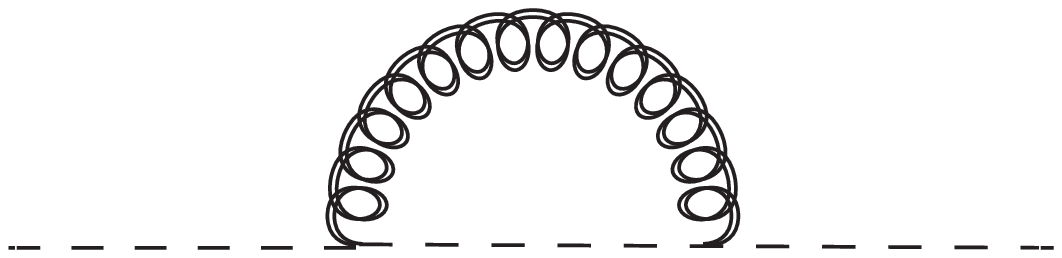}
\end{tabular}& $+$  &\begin{tabular}[c]{c}
\includegraphics[scale=0.4]{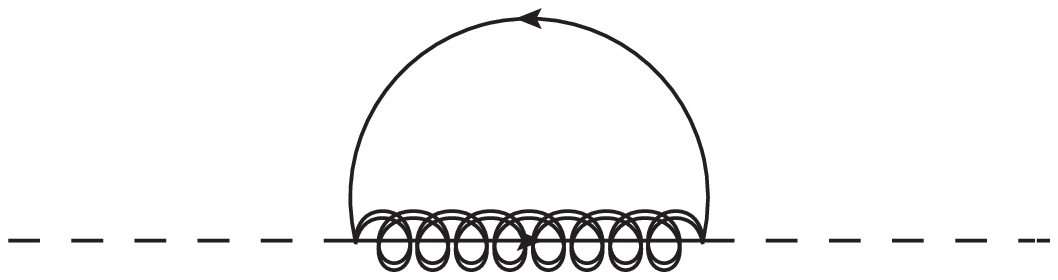}
\end{tabular}\\
& \begin{tabular}[c]{c}
\includegraphics[scale=0.4]{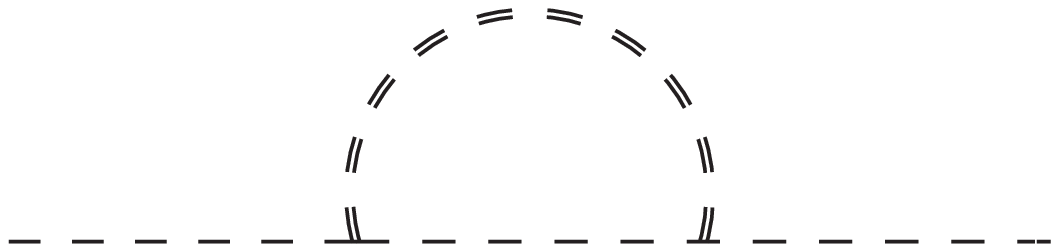}
\end{tabular}
\end{tabular}
\end{center}
\caption{Where the external legs correspond to the MSSM squarks, and the doubled wavy-solid (wavy) lines are Higgsed gauginos (gauge bosons), double dashed lines are the scalar messengers. The leading F/M contribution is given by the diagram with the Higgsed gaugino (with 4 mass insertions).}
\label{oneloopscalarmass}
\end{figure}

Scalar masses can be generated at one loop. This is because the gauge messengers couple to non-zero F-terms already at tree-level. The diagrams are presented in Figure \ref{oneloopscalarmass}.

The crucial observation is that in the SUSY limit the mass of the Higgsed vector superfields is given by $(M^2_v)^{AB}=\Phi^{\dagger}\{T^{A},T^{B}\}\Phi$. One can take the simplifying assumption that all the masses are the same and that we've chosen a basis where they are diagonal, i.e. $(M^2_v)^{AB}=(\Phi,\Phi)\delta^{AB}$, where the inner product is defined as the (degenerate) eigenvalue of the matrix $\Phi^{\dagger}\{T^{A},T^{B}\}\Phi$.

The one loop R.G. equation for the Quark superfield is given by:
\begin{equation}
 \frac{d}{dt}Log(Z_Q)=\frac{c}{\pi}\alpha
\end{equation}
where $c'$ is the Casimir of the Quark superfield representation under GUT gauge group ($c'=\frac{N^2-1}{2N}$ for an $SU(N)$ fundamental).

So that below the messenger threshold the wave-function renormalization function is given by:
\begin{equation}
 Z_Q(M_v,\mu)=Z_Q(\Lambda_{U.V}\left(\frac{\alpha(\Lambda_{U.V.})}{\alpha(M_v)}\right)^{\frac{2c'}{b'}} \left(\frac{\alpha(\Lambda_{M_v})}{\alpha(\mu)}\right)^{\frac{2c_a}{b_a}}
\label{wavefunctionrenormalization}
\end{equation}
and $b'$ is the beta-function coefficient of the GUT gauge coupling and c,b are the corresponding constants for the MSSM gauge couplings.

The gauge coupling below the messenger threshold is given by:
\begin{equation}
 \alpha^{-1}(\mu)=\alpha^{-1}(\Lambda_{U.V.})+\frac{b'}{4\pi}Log(\frac{(X,X)}{\Lambda^2_{U.V.}})+\frac{b}{4\pi}Log(\frac{\mu^2}{(X,X)})
\label{gaugecoupling}
\end{equation}
We now need to continue these expressions into superspace and extract the SUSY breaking soft terms. The first step is to canonically normalize the fields:
upon analytically continuation Z picks up $\theta^2$,$\overline{\theta}^2$ and $\theta^2\overline{\theta}^2$ terms with the constraint that it must be a real function. So the $\theta^2$ and the $\overline{\theta}^2$ components are the complex conjugates of each other:
\begin{equation}
 Z=z+Z|_{\theta^2}\theta^2+(Z|_{\theta^2})^{\dagger}\overline{\theta^2}+Z|_{\theta^2\overline{\theta}^2}\theta^2\overline{\theta}^2
\end{equation}
Where the z is the ``scalar`` component of Z, $Z|_{\theta^2}$ is the $\theta^2$ component of Z and $Z|_{\theta^2\overline{\theta}^2}$ is the $\theta \overline{\theta}^2$ component of Z.

Then after the redefinition of canonically normalized fields:
\begin{equation}
 Q'=z^{1/2}(1+\frac{Z|_{\theta^2}}{z}\theta^2)Q
\end{equation}
Where Q is the normal Quark superfield and Q' is the canonically normalized quark superfield.

If we define the terms in the potential as:
\begin{equation}
 V=\sum_i m^2_{Q_i}Q_i^{\dagger}Q_i+A_{Q_i} Qi\frac{\partial W}{\partial Qi}+c.c. +(\frac{\partial W}{\partial Qi})^{*}\frac{\partial W}{\partial Qi}
\end{equation}
And integrate out the auxiliary components of the quark fields using the previous expressions, if follows that:
\begin{eqnarray}
A_{Q_i}=Log(Z_{Q_i})|_{\theta^2}\\
m^2_{Q_i}=-Log(Z_Q)|_{\theta^2 \overline{\theta}^2}
\label{Atermsandsquarkmasses}
\end{eqnarray}
Because the correct mass threshold for gauge messengers is of the form $(x,x)$ and not $X^{\dagger}X$, the expansion of the gauge coupling into superspace is given by:
\begin{eqnarray}
 \alpha^{-1}((X,X))=\alpha^{-1}((x,x))+\frac{b'}{4\pi}(\theta^2 \frac{(x,F)}{(x,x)}+\overline{\theta}^2 \frac{(F,x)}{(x,x)}+\theta^2 \overline{\theta}^2 \frac{(F,F)(x,x)-(x,F)(F,x)}{(x,x)^2})\\
\alpha_a^{-1}(\mu)=\alpha_a^{-1}(\mu)+\frac{b'-b_a}{4\pi}(\theta^2 \frac{(x,F)}{(x,x)}+\overline{\theta}^2 \frac{(F,x)}{(x,x)}+\theta^2 \overline{\theta}^2 \frac{(F,F)(x,x)-(x,F)(F,x)}{(x,x)^2})
\end{eqnarray}
Where $\alpha_a^{-1}(\mu)$ is the gauge coupling below the gauge messenger mass threshold and $\alpha^{-1}((X,X))$ is the gauge coupling at the messenger mass threshold.

If we now replace these expressions for the gauge couplings in the expression for the squark masses, we get:
\begin{align}
 m^2_Q= &\frac{g^2(\mu)}{8\pi^2}((c-\chi c')+c \frac{b'}{b}(\chi-1))\frac{(F.F)(x,x)-(x,F)(F,x)}{(x,x)^2}+\nonumber\\
&+2\left(\frac{g^2(\mu)}{16\pi^2}\right)^2((bc+b'(c'-2c))+(\chi^2-1)\frac{b'}{b}(bc'-cb'))\frac{(x,F)(F,x)}{(x,x)^2}
\label{squarkmasses}
\end{align}
Where $\chi=\frac{\alpha(M)}{\alpha(\mu)}$. Where we note that the one loop contribution is often tachyonic since $c \le c'$ and $(F.F)(x,x)-(x,F)(F,x)\ge 0$.

The case with two messenger thresholds is a simple generalization of this, and at leading order (i.e. ignoring the running of the gauge coupling) the result is a direct sum of the result we got for gauge messengers and the usual result for normal messengers. So the contributions for the soft terms coming from different messengers add up\footnote{See appendix \ref{RGtwothresholds} for more details}.

\subsection{A-terms:}
The relevant diagram to compute is:

\begin{figure}[h!]
\begin{center}
 \begin{tabular}{cc}
$A \approx$ &
\begin{tabular}[c]{c}
\includegraphics[scale=0.4]{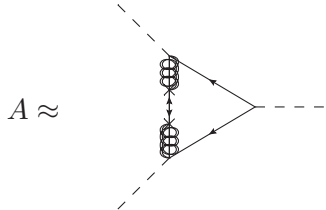}
\end{tabular}
\end{tabular}

\caption{Main diagram contributing to the trilinear couplings at leading order in F/M.}
\label{trilinearcouplingsdiagram}
\end{center}
\end{figure}
As we saw in the previous section, the A-terms are given by:
\begin{equation}
A_Q=Log(Z_Q)|_{\theta^2}
\end{equation}
Being that this expression is readily extracted from considering the expression for the wave-function renormalization: (\ref{wavefunctionrenormalization}), and the expression for the gauge coupling ($\ref{gaugecoupling}$) analytically continued to superspace.

The result is given by:
\begin{equation}
 A(\mu)=\frac{\alpha(\mu)}{2\pi} (c'-c) \frac{(x,F)}{(x,x)}+\frac{\alpha(\mu)}{2\pi} (c'-c\frac{b'}{b})(\chi-1) \frac{(x,F)}{(x,x)}
\end{equation}
Where $\chi=\frac{\alpha(M)}{\alpha(\mu)}$.

so that the main contribution is
\begin{equation}
 A(\mu)\approx\frac{\alpha(\mu)}{2\pi} (c'-c) \frac{(x,F)}{(x,x)}
\end{equation}

\subsection{Suppression of the one loop contribution to squark masses:}

An important result of \cite{Intriligator:2010be} was to show that when the scalar partner of the goldstino (sgoldstino) gets a large vev, these corrections are suppressed.

It is a well known result that in O'R. models where the superpotential is a cubic polynomial in the fields, the vev of the scalar partner of the goldstino parameterizes a flat direction of the potential\cite{Komargodski:2009jf,Ray:2006wk}:
\begin{equation}
 x\rightarrow x+ z F
 \label{flatdirection}
\end{equation}
is a flat direction.

The one loop contribution is proportional to the coefficient$\frac{(F,F) (x,x)-(x,F)(F,x)}{(x,x)^2}$, which along the flat direction scales as:
\begin{align}
 &\frac{(F,F) (x+z F,x+z F)-(x+zF ,F)(F,x+z F)}{(x+z F,x+z F)^2}= \nonumber\\
&\frac{(F,F) (x,x)-(x,F)(F,x)}{(x+z F,x+z F)^2}\rightarrow\frac{(F,F) (x,x)-(x,F)(F,x)}{|z|^4 (F,F)^2}
\end{align}
Where we took the limit $zF\gg x$. Comparing with the suppression one gets for the usual two loop contribution:
\begin{equation}
 \frac{(F,x+z F)(x+z F,F)}{(x+zF,x+zF)^2}\rightarrow \frac{1}{|z|^2}
\end{equation}
So that the ratio between the one and the two loop contribution is:
\begin{equation}
 \frac{(m^2_Q)^{(1)}}{(m^2_Q)^{(2)}}\approx \frac{4\pi)}{\alpha(\mu)}\frac{c'-c''}{(b'c'+b''(c''-2c')}\frac{(F,F) (x,x)-(x,F)(F,x)}{(F,F)^2}\frac{1}{|z|^2}
\end{equation}
so that if we stabilize z sufficiently far away from the origin the two loop corrections can dominate over the one-loop ones.
\begin{equation}
 z\ge \sqrt{\frac{4\pi}{\alpha(\mu)}\frac{c'-c''}{(b'c'+b''(c''-2c'))}\frac{(F,F) (x,x)-(x,F)(F,x)}{(F,F)^2}}
\end{equation}
We note that since $(F,F) (x,x)-(x,F)(F,x)$ can be small when compared with $(F,F)$ due to some alignement, one can have a sufficient suppression of the one loop correction without necessarily requiring a very large value of z.

If instead of z we had used $y=\frac{z}{|F|}$ to parameterize the flat direction, the limit is approximately given by:
\begin{equation}
 y > \frac{4\pi}{g^2}M_v
\label{oneloopbound}
\end{equation}
Where $M_v$ is the mass of the Higgsed gauginos evaluated at the origin of the pseudomoduli space $y=0$ (If there is some alignment between F and x this lower bound can be violated).

\section{Model Building Constraints:}

In this section we will show that in a large class of models, for large sgoldstino vevs, one can show that for every messenger in a model:
\begin{equation}
 \frac{F}{M} \approx \frac{1}{z}
\end{equation}
Where the pseudomoduli direction is given by: $X_i=x_i^{(0)}+z F_i$\footnote{Since the Kahler potential is canonical, in the vacuum, $F_i=W_i$, so both quantities can be used}. What this means is that in a large class of models, $F/M$ is the same for all (gauge and non-gauge) messengers. In other words: even if the mass thresholds of the messengers are different and the F-terms they couple to are different, $F/M$ will be the same for all messengers. What this implies is that the sign of the squark/slepton masses depends only on the field content of the theory, not on it's parameters \footnote{The overall scale, i.e. the value at which z is stabilized, will of course depend on the parameters of the model. Also, if the mass splittings are too large R.G. effects should be taken into account.}.

Let us be more precise about what kind of models we are considering:
\begin{itemize}
 \item The Kahler potential is canonical;
 \item SUSY should be broken at tree-level (no runaway directions);
 \item The vev of the sgoldstino parameterizes the only flat direction and it should be everywhere stable (i.e. no metastability);
 \item The messengers couple linearly to the goldstino;
 \item The fermionic mass matrices for (non-gauge) messengers factorize into $2\times 2$ matrices;
 \item In order to suppress the one loop tachyonic contribution from gauge messenger to scalar masses the sgoldstino is the largest vev in the model;
\end{itemize}

For example: the first four constraints are easily satisfied in renormalizable theories that break SUSY by virtue of the rank condition (and we sit at the global minima). The fifth condition essentially tells us that the messenger sector should not contain three fields that can mix in complicated ways.

An important result \cite{Komargodski:2009jf,Shih:2007av} is that in the global minima one has:
\begin{equation}
 \frac{\partial}{\partial X}Det(W_{ij})=0
\label{vanishinggaugino}
\end{equation}
Where X is the vev of the scalar partner of the goldstino, and $W_{ij}$ is the second derivative of the superpotential with respect to the fields $\phi_{i},\phi_{j}$. This is nothing but the argument that gaugino masses vanish at leading order in $F/M$ unless the vacuum is metastable.

We also note that if the superpotential for the messengers can be written in the form, and there is an R-symmetry:
\begin{equation}
 W=f X+(M^{ij}+X N^{ij})\phi^{i}\phi^{j}
\end{equation}
then\cite{Shih:2007av}:
\begin{equation}
 Det(M+X N)=Det(M)
\end{equation}
wether or not there is metastability, so that $\frac{\partial}{\partial X}Det(W_{ij})=0$, and the following argument still applies.

By manipulating (\ref{vanishinggaugino}), and choosing a basis where the fermionic mass matrix is diagonal, one can rewrite is as:
\begin{equation}
 Tr(\frac{W_{ijk}F^{k}}{m_i})=0
\end{equation}
And $W_{ij}=m_i \delta_{ij}$.

When the fermionic mass matrix factorizes to a $2\times 2$ matrix, this means that:
\begin{equation}
 \frac{W_{11k}F^{k}}{m_1}=-\frac{W_{22k}F^{k}}{m_2}
\label{equalityofFM}
\end{equation}
So, the contribution to the soft scalar masses is exactly the same for both fermionic mass eigenstates (at leading order in F/M).

We now write down the general dependence of the mass matrix for the messengers on the vevs of the model:
\begin{equation}
 W_{ij}=m_{ij}+W_{ijk}X^{k}+W_{ijl}\phi^{l}+O(\phi^2)
\end{equation}
Where $m_{ij}$ is some mass matrix and $W_{ijk}$ is the third derivative of the superpotential, and we have separated the dependence on the goldstino field X from the other fields ($\phi$'s). The higher order terms are absent if the superpotential is a cubic polynomial in the fields. Since messengers couple to the goldstino, the term $W_{ijk}X^{k}$ cannot be identically 0.

For large vevs of X\footnote{Note that (\ref{oneloopbound}) gives a bound on how big the other (non-gauge invariant) vevs can be in order to have an appropriate suppression of the one loop contributions.} we have:
\begin{equation}
 Tr(W_{ij}) \approx W_{iik}X^{k}
\end{equation}
This means that:
\begin{equation}
\begin{array}{c}
Tr(W_{ij})=W_{iik}X^{k}\\
Det(W_{ij})=constant
\end{array}
\end{equation}
For the mass matrices we are considering, this implies that one of the mass eigenstates is very light and the other is very heavy:
\begin{equation}
\begin{array}{c}
m_H\approx W_{iik}X^{k}\\
m_L\approx \frac{constant}{W_{iik}X^{k}}
\end{array}
\end{equation}
So, for the heavy field, the contribution to the soft masses is given by:
\begin{equation}
 \frac{W_{iik}F^{k}}{m_i}\approx \frac{W_{iik}F^{k}}{W_{iik}X^{k}}=\frac{W_{iik}F^{k}}{W_{iik}(x^{k,(0)}+z F^{k})}\approx\frac{W_{iik}F^{k}}{z W_{iik}F^{k}}= \frac{1}{z}
\end{equation}
And (\ref{equalityofFM}) tells us that this contribution is the same for both mass eigenstates.

Now, for the gauge messengers, we've just shown that the contribution to soft masses is proportional to the square of $\frac{(F,X)_i}{(X,X)_i}$, where the inner product $(A,B)_i$ is defined as the i-th eigenvalue of $A^{\dagger}\{T^a,T^b\}B$.

For large vevs of X this is given by:
\begin{equation}
\frac{(F,X)}{(X,X)}=\frac{(F,x+z F)}{(x+z F,z+z F)} \approx \frac{z(F,F)}{z^2(F,F)}=\frac{1}{z}
\end{equation}
As we wanted to show.

\subsection{Constraining models with gauge messengers:}

We've seen that in a variety of models the ratios $F/M$ for the different fields approach a universal value for large values of the sgoldstino vev. This means that the sign of the squark and slepton masses generated is a function only of group theory factors (i.e. the representations of the messenger fields) and does not depend on the superpotential parameters. In particular, apart from R.G. effects, the ratios of squarks and sleptons depend only on the representations of the messenger fields.

It is interesting to note that in these models, for large sgoldstino vevs, we recover the condition for non-tachyonic scalar soft masses that was derived in \cite{Giudice:1997ni}:
\begin{equation}
 b'(2-\frac{c'}{c_i})<b_i
\label{boundonmessengers}
\end{equation}
Where $b'$ $(b_i)$ are coefficients of the beta functions for the gauge couplings above (below) the GUT scale and $c'$ ($c_i$) are the Casimirs of the representation under which the field transforms above (below) the GUT scale.

The main difference though is that equality of $F/M$ for all messengers is not assumed.

This constraint can be easily evaded if we assume metastability. There are different reasons why this scenario is probably not preferred. Metastability usually requires the existence of an approximate R-symmetry \cite{Nelson:1993nf,Intriligator:2007py,Murayama:2006yf,Murayama:2007fe,Kitano:2006xg}. Unlike in normal models of gauge mediation (e.g. ISS\cite{Intriligator:2007py}) to suppress the tachyonic contribution to squarks we do not want the vacuum to be close to the origin of field space. It is hard to see how a polynomial superpotential could have an approximate R-symmetry far away from the origin.

 If the gauge coupling (above the GUT scale) is asymptotically free, the mechanism of \cite{Witten:1981kv} becomes available to stabilize the vacuum far away from the origin. However, this doesn't always happen. Also, the one loop contributions to the effective potential don't necessarily lead to the stabilization of flat directions\cite{Dine:2007dz,Matos:2009xv,Dine:2006xt,Intriligator:2010be}.

Another thing to consider is wether such local minima should be preferred with respect to global ones. In \cite{Abel:2006my,Abel:2006cr,Craig:2006kx,Fischler:2006xh} it was shown that generically thermal corrections in the early universe make vacua close to the origin of field space preferred, this is because close to the origin of field space there are usually more light fields, so thermal corrections are smaller.

It is also possible that the number of messengers is sufficiently small so that (\ref{boundonmessengers}) is verified. However, two adjoints is already too many...

One could think of building a model with one adjoint field (plus messengers in other representations), but this is very hard. Assume:
\begin{itemize}
 \item There is only one adjoint field Y (plus fields in other representations);
 \item In order to break $SU(5) \rightarrow SU(3)\times SU(2) \times U(1)$, Y is the only non-gauge invariant field whose F-term and scalar component can get vevs (singlets can have scalar and F-term vevs);
\item Above the GUT scale, the model is SUSY;
\item The sgolstino vev parameterizes a flat direction;
\end{itemize}
The second constraint is actually not very constraining as fundamental, symmetric and anti-symmetric vevs don't break $SU(5)$ to the MSSM gauge groups. More complicated representations are likely to contribute to the beta function so that (\ref{boundonmessengers}) is violated.

The third condition means that no field that couples to any SUSY breaking vev can be heavier than the GUT scale\footnote{We are ignoring the possibility of non-zero D-terms.}. In other words: the only SUSY breaking parameters allowed at the GUT scale are F-term vevs of dynamical fields, there are no spurions.

In order to have gauge messengers, there must exist a non-gauge invariant F-term. This means that in the vacuum:
\begin{equation}
 Y=y+\theta^2 F_{y}
\end{equation}
And these vevs break $SU(5) \rightarrow SU(3) \times SU(2) \times U(1)$.

Since there is only one adjoint,it must be that:
\begin{equation}
 F_{y} \propto f(y)
\end{equation}
Since y is the only non-gauge invariant vev. If we assume the superpotential is a polynomial function of the fields, we have that $f(0)=0$.

However in models where SUSY is broken due to the rank condition, or where SUSY is spontaneously broken and the superpotential is a cubic polynomial, the vev of the goldstino parameterizes a flat direction. Since $F_{y}$ is non-zero, Y is part of the goldstino (generally there may be more non-zero F-terms, so that the goldstino is a particular linear combination of these fields).
In any case, the scalar component of Y, y, parameterizes a flat direction. So $F_{y}$ cannot depend on y and must vanish.

So, with these constraints it's not possible to build a model of gauge mediation where gauge messengers exist.

Another possible way around this problem which we shall not explore here is to embed the $SU(5)$ into a product of several groups \cite{Dimopoulos:1997ww,Murayama:1997pb}.

\section{Possible solutions:}

A simple possible solution to this problem is to have a second independent flat direction that does not couple to (all) normal messengers. This violates one of the constraints and allows us to address the problem of tachyonic scalar masses.

We will now consider two ways in which this can happen:

\begin{figure}[h!]
\begin{center}
\begin{tabular}{cc}
Case 1 & Case 2\\
\includegraphics[scale=0.25]{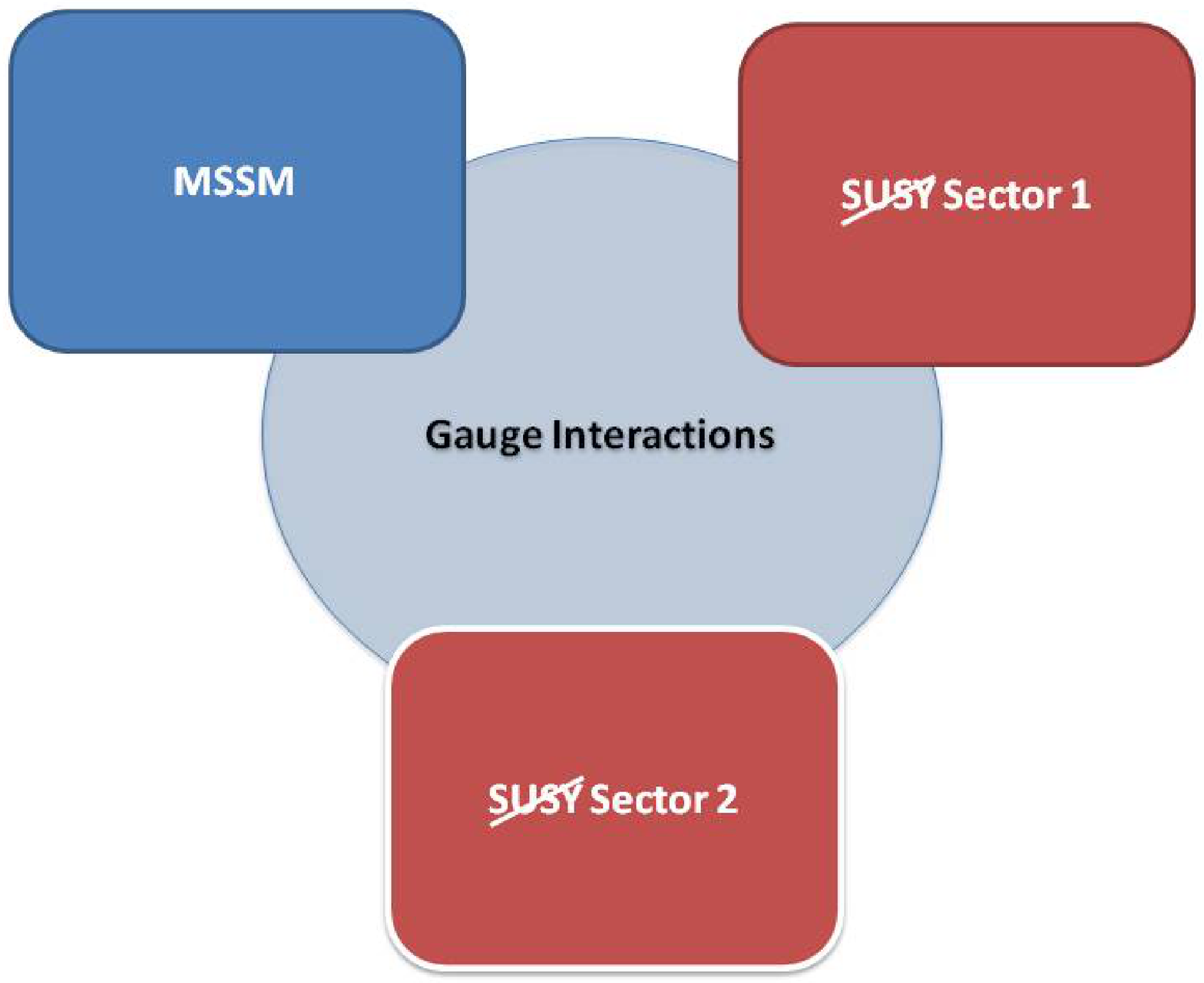} & \includegraphics[scale=0.25]{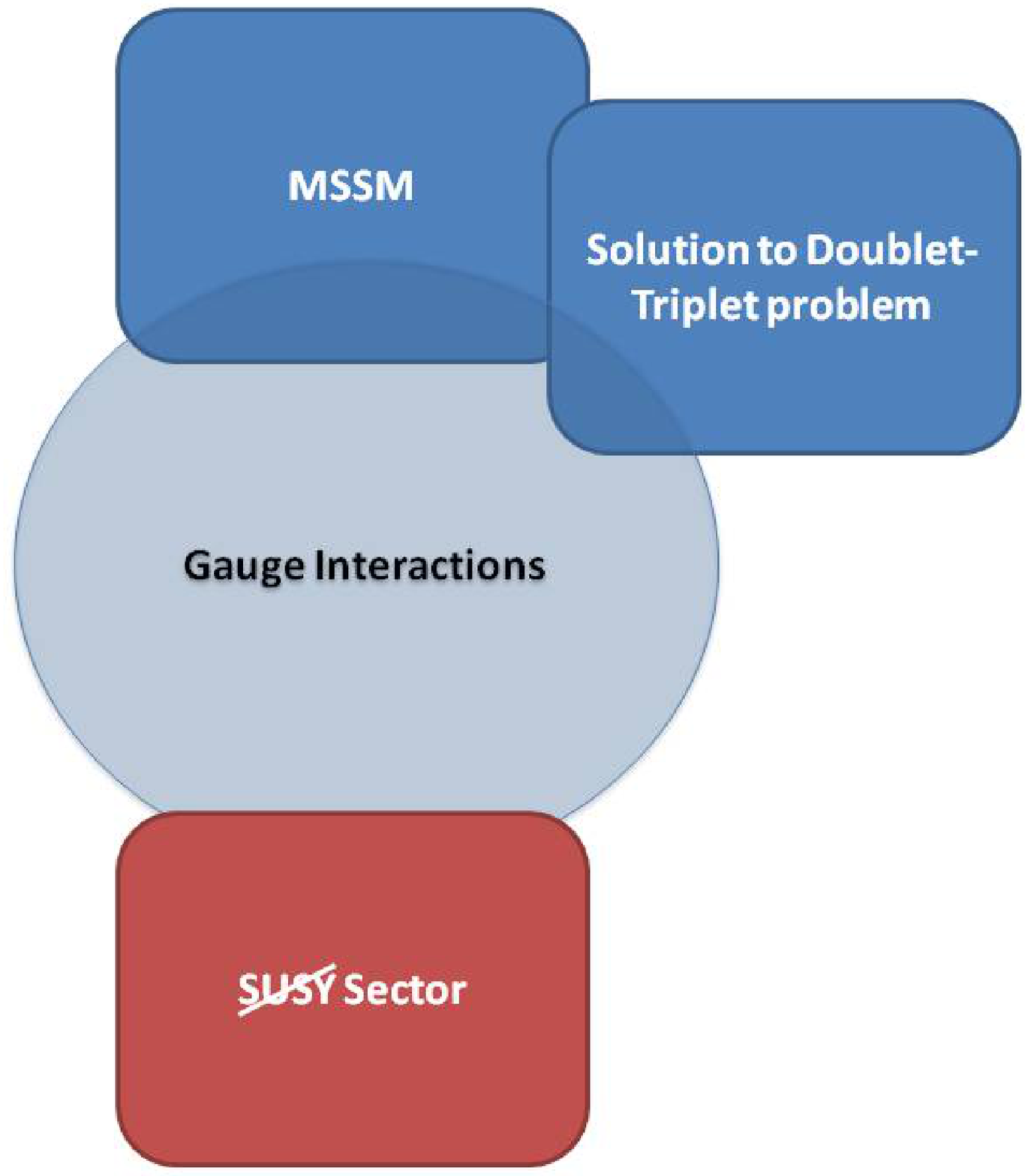}
\end{tabular}
\label{pictoricalrepresentation}
\end{center}
\end{figure}

Where it's assumed that the solution to the doublet-triplet problem implies the existence of an adjoint field whose vev breaks $SU(5)$ to the MSSM gauge groups, or some other field whose vev contributes to the mass of the Higgsed vector superfields. Related studies to the second case have been done in \cite{Nardecchia:2009ew}, but where the emphasis is direct gauge mediation, and in \cite{FileviezPerez:2009im}, where there are similar couplings between an extra sector that contains an adjoint. They solve the tachyon problem by simply coupling the adjoint to a spurion singlet field through superpotential interactions, so the messengers are only chiral fields.

\subsection{Case 1:}

In this case we create the second flat direction by adding a second sector where SUSY is spontaneously broken and that does not contain gauge messengers. There are then two goldstino fields. By changing the ratio of the vevs of the two sgoldstino fields we can enhance the contributions from normal messengers to the soft terms, and get non-tachyonic squarks and sleptons.

As an example consider that we couple the model we presented in section 1, and add a second sector with 2 chiral messengers in fundamental/anti-fundamental pairs ($Q_1,Q_2,\tilde{Q}_1,\tilde{Q}_2$). The superpotential is:
\begin{equation}
  W=-\mu^2 \Phi+\lambda \Phi Tr(Y_0 Y_0)+m Tr(Y_2.Y0)+\overline{\lambda}Tr(Y_2 Y_0 Y_0)+\lambda_2 \overline{\Phi} Q_1.\tilde{Q}_1-\overline{\mu}^2\overline{\Phi}+m1 Q_2 \tilde{Q}_1+m_2 Q_1 \tilde{Q}_2
\end{equation}
Where $\Phi,\overline{\Phi},Q_2,\tilde{Q}_2$ have R-charges equal to 2, and $Y_0, Q_1,\tilde{Q}_1$ have R-charge 0. Since there are no couplings between the fields of the two sector, we can study them independently.

To simplify the discussion even further, we shall take $m_1=m_2$ and $\overline{\mu}=\mu$. For $m_2^2>\lambda_2 \mu^2$, the quarks do not get vevs, and the minimum of the potential is given by:
\begin{equation}
\begin{array}{c}
Q_1=0;\
Q_2=0;\\
\tilde{Q}_1=0;\
\tilde{Q}_2=0;\\
\end{array}
\end{equation}
And $\overline{\Phi}$ is undetermined at tree-level.

 Furthermore, since in this sector there are no gauge messengers \cite{Shih:2007av}, quantum corrections stabilize the vev of the goldstino at the origin of field space. It is not relevant that R-symmetry is not spontaneously broken in this sector, what is important is that the sgolstino vev of the sector with gauge messengers is non-zero (as this breaks the R-symmetry). At the origin of field space, both quarks get masses equal to m, and couple to an F-term equal to $\lambda \mu^2$, so that
\begin{equation}
 \frac{F}{M}_{n.s.}=\frac{\lambda \mu^2}{m_2}
\end{equation}
Where this contribution only affects soft scalar masses, and n.s. stands for ``normal messenger sector``.

We have already studied the other sector, and for large sgoldstino vevs F/M is approximately given by:
\begin{equation}
 \frac{F}{M}_{g.s.}=\frac{4 \lambda \mu^2}{\overline{\lambda} y}
\end{equation}
Where g.s stands for gauge messenger sector and y parameterizes the vev along the sgoldstino direction. These contributions affect both soft gaugino and scalar masses. In particular the contribution to the scalar masses is negative.

We will now specify the region in parameter space that we will be considering. In order to keep the gauge couplings in the perturbative regime, all the particles reasonably heavy. Also, to keep the scale of the soft terms much lower than the GUT scale to avoid a large tuning for electroweak symmetry breaking, $\sqrt{F}\ll M_{GUT}$. There are several ways to do this. The way we will do it is by choosing the parameters around the region where the F-term equations become degenerate and SUSY stops being broken, i.e. if the superpotential was:
\begin{equation}
 W= X_1 f_1(\phi)+X_2 f_2(\phi)
\end{equation}
We would choose the parameters in such a way that the two $f_i(\{\phi_j\})$'s vanish at the same point in field space. A possible reason for this to happen could be an approximate symmetry of the superpotential of the high energy theory (e.g. it is only violated by some non-perturbative term).

In this work we will assume that the flat direction can be stabilized far away from the origin. In \cite{Dine:2006xt} it was shown that in models with gauge messengers, even if the R-charges of the fields are 2 and 0 the sgolstino vev can be stabilized away from the origin. So, Coleman-Weinberg corrections are a possible mechanism to achieve this.

The scalar two loop contributions are given by:
\begin{equation}
 m_q^2 \approx \frac{\alpha(\mu)^2}{8\pi^2}(N c'_i (\frac{F}{M}_{n.s.})^2+((c''-2c'_i)b''+b'_ic')(\frac{F}{M}_{g.s.})^2)
\end{equation}
where $c'',c'_i$ are the quadratic Casimirs for the particular MSSM quark (for a fundamental of $SU(5)$, $c''=\frac{12}{5}$, and $c'_3=\frac{4}{3},c'_2=\frac{3}{4}$) and N is the index of the matter messengers, for this model $b''=-4,b'_3=1$, and $b'_2=-3$, $N=2$.
(gaugino and trilinear couplings are both non-zero at leading order in F/M and given by the respective expressions)

and we can choose the parameters in such a way that the vev of the field y is such that all the squarks/sleptons are non-tachyonic.

We will now give an example point:
\begin{table}[!h]
\begin{center}
 \begin{tabular}{|c|c|c|c|c|c|c|c|}
 \hline
$\mu$ & $\lambda$ & $\lambda_2$ & $\lambda_3$ & $\frac{\mu_2}{\mu}$     & $\frac{m}{\mu}$   & $\frac{\overline{m}}{\mu}$    & $g_{GUT}$\\[2mm]
 \hline
 \hline
 $1$  &    $0.5$  &     $0.5$   &     $0.1$   &   $1.1\times 10^{-4}$   & $0.1291$          &      $ 1$            &  $  1$\tabularnewline
 \hline
 \end{tabular}
\end{center}
\end{table}
In units of $\mu$. We assume that the goldstino flat direction in the gauge messenger is stabilized for $y=23.85$, and for the non-gauge messenger sector is stabilized at the origin. Computing the mass of the Higgsed vector fields, allows us to match $\mu$ in units of $M_{GUT}$: $\mu=0.022M_{GUT}$.

The masses of the other fermionic fields are (in units of $M_{GUT}$):
\begin{table}[!h]
\begin{center}
 \begin{tabular}{|c||c|}
 \hline
Adjoints $SU(3)$ & $(0.5,0.0002)$\\
 \hline
Adjoints $SU(2)$ & $(0.5,0.0002)$\\
 \hline
Fundamentals  & $0.017$\\
 \hline
Bifundamentals  & $(1,1,1.92 \times 10^{-9})$ \\
 \hline
 \end{tabular}
\end{center}
\end{table}

The ``effective'' F/M one loop contribution is $\frac{F}{M}^{(1)}=\sqrt{\frac{(F.F)(x,x)-(x,F)(F,x)}{(x,x)^2}}=4 \times 10^{-13}M_{GUT}$.

The two loop $F/M$ contribution is:
\begin{equation}
 \begin{array}{c}
(F/M)_{g.s.}=9,00\times 10^{-13} M_{GUT}\\
(F/M)_{n.s.}=1,03\times 10^{-11} M_{GUT}
 \end{array}
\end{equation}
Where in the sector with the gauge messengers, all $F/M$'s are approximately the equal to the value $F/M_{g.m.}$, and for the sector with the quarks $F/M$ is given by $F/M_{fund}$.

We can now compute the squark and slepton masses, which we summarize in the next table:
\begin{table}[!h]
\begin{center}
\small{ \begin{tabular}{|c|c|c|c|c|c|c|}
\hline
Field           &   $Q$                  & $U$                              &  $D^{c}$                        &  $L$                  & $E^{c}$           & $H_u$ \\
\hline
\hline
$\frac{m_Q}{M_{GUT}}$          &   $3.53 \times 10^{-13}$ & $2.96 \times 10^{-13}$  & $2.29 \times 10^{-13}$ & $2.60 \times 10^{-13}$ & $3.16 \times 10^{-13}$ & $2.61 \times 10^{-13}$\\
\hline
 \end{tabular}}
\end{center}
\caption{Squark and sfermion soft scalar masses.}
\end{table}
Where these soft masses are computed at the messengers scale (i.e. close to the GUT scale) and we took all gauge coupling to be equal, and equal to 1.

A-terms are proportional to:
\begin{equation}
 A\approx \frac{g^2}{8\pi^2}(c-c')\frac{(F,x)}{(x,x)}\approx 3 \times 10^{-14}M_{GUT}
\end{equation}
While the gaugino masses are around:
\begin{equation}
 m_{\lambda} \approx- \frac{g^2}{8\pi^2}(5-N_i)\frac{(F,x)}{(x,x)}\approx -4 \times 10^{-14}M_{GUT}
\end{equation}
If $M_{GUT}$ is take to be $10^{16}$, then squarks and sleptons have masses around 3TeV and gauginos have masses around 400GeV, at the messenger scale.

More generally, and in the worst case, one may expect an approximate upper bound on the ratio:
\begin{equation}
 \frac{m_{\lambda}}{m_Q}\lesssim \frac{1}{4\pi}
\end{equation}
So that even if the sgolstino of the sector with gauge messengers is not stabilized very far away from the origin, the splitting between gauginos and squarks/sleptons is around one order of magnitude.

In this model one does not expect unification to be automatic. At one loop, and with the parameters we used, it is very simple to calculate the gauge couplings as a function of the energy scale:
\begin{figure}[!h]
\begin{center}
\includegraphics[scale=0.9]{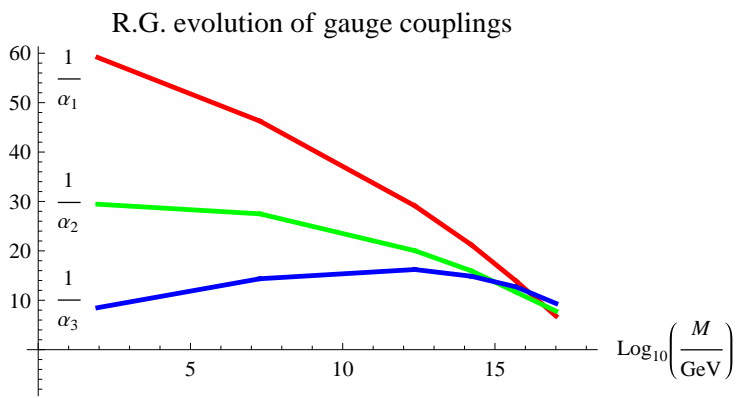}
\caption{One Loop R.G. equations for the Gauge Couplings.}
\label{RGexample1}
\end{center}
\end{figure}

As we can see in Figure \ref{RGexample1}, unification is possible without a large fine-tuning (but is not automatic).

 \subsection{Case 2:}

In this case there is a flat direction to which the gauge messengers couple, but the normal messengers do not. Unlike in the previous example this flat direction is not associated with SUSY breaking. So the normal messengers have masses of order $m=X$ (where X is the sgoldstino vev), and gauge messengers have masses of order $\Phi$, where $\Phi$ is the vev along the second flat direction. In this case the suppression of the one loop tachyonic contribution of the gauge messengers to the soft squark/slepton masses is lost.

The expressions of the soft terms is approximately:
\begin{equation}
 \begin{array}{c}
m_{Q}^2 \approx -(\frac{g^2}{16\pi^2}) N_{g.m.}(\frac{F}{\Phi})^2+(\frac{g^2}{16\pi^2})^2N_{n.m.}(\frac{F}{X})^2\\
m_{\lambda} \approx \frac{g^2}{16\pi^2}\overline{N}_{g.m.}\frac{(F,X)}{(\Phi,\Phi)}
 \end{array}
\end{equation}
Where the group theory factors associated with the number of messengers are encoded in $N_{g.m.},N_{n.m.}$ and $\overline{N}_{g.m.}$. Where X is the sgoldstino vev and $\Phi$ is the vev along the flat direction. One then needs $\frac{\Phi}{X}$ to be large enough so that $m_{Q}^2>0$.

If $\frac{\Phi}{X}$ is large enough so that $m_{Q}^2>0$, a significant cancelation between the one loop contribution from gauge messengers and the two loop contribution from normal messengers is required to get $m_{Q} \sim m_{\lambda}$.

 \subsubsection{Example:}
As a particular example, we now explore the possibility that this solution is actually connected to the solution of the doublet-triplet problem. We will now briefly review this problem\cite{Raby:2006sk,Raby:2008gh}.

In the context of $SU(5)$ GUTs the doublet-triplet problem can be understood in the following way: take the two MSSM Higgs fields to be a fundamental/antifundamental pair of $SU(5)$. Below the energy at which the GUT symmetry is spontaneously broken these representations split to two $(3,1)$ and two $(1,2)$ under $SU(3)$ and $SU(2)$ respectively. The $(1,2)$'s are the MSSM Higgs fields. Since the triplets are absent in the low energy theory, they must be massive. Below these Higgs triplets mass scale, they can be integrated out. This generically generates dimension 5 operators\footnote{Strictly speaking the dimension 5 operators that we are considering are the effective vertices one gets from integrating out the Higgs triplet and Higgsed gaugino (i.e. Consider that the Higgs triplet and Higgsed gaugino propagators are evaluated at zero momenta and contracted to a point). For clarity we present the full diagram.} that allow for proton decay\cite{Dimopoulos:1981dw}:

\begin{figure}[h!]
\begin{center}
 \begin{tabular}{c}
\includegraphics[scale=0.4]{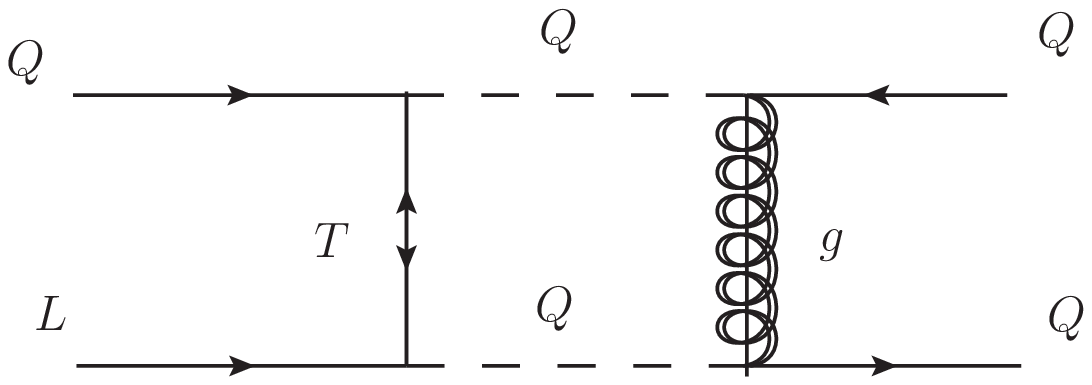}
\end{tabular}
\end{center}
\end{figure}
Where T represents the Higgs triplets, Q and L are quarks/leptons (for solid lines) and squarks/sleptons (for dashed lines), and we suppressed the family and gauge indices. This diagram is suppressed by the Higgsino triplet mass and the SUSY breaking scale. Since there are very stringent bound on this decay, it means that the Higgs triplets should actually be very heavy.

This problem can be addressed if one assumes the existence of a sliding singlet that couples to the Higgs \cite{Ibanez:1982fr}:
\begin{equation}
 W=\lambda_x (H_u.\tilde{Y}_0.H_d+\tilde{\Phi} H_u.H_d) +m_2 Tr[\tilde{Y}_0.\tilde{Y}_0]+\overline{\lambda}_{x} Tr[\tilde{Y}_0.\tilde{Y}_0.\tilde{Y}_0]
\end{equation}
This superpotential has many different minima, all of which are SUSY. If all the Higgs vevs are 0, we have:
\begin{equation}
 \begin{array}{ccc}
  H_u=0 & H_d=0 & \tilde{Y}_0=0\vspace{15pt}\\
  H_u=0 & H_d=0 & \tilde{Y}_0=\frac{2 m_2}{3 \overline{\lambda}_{x}}diag(\{2,2,2,-3,-3\})\\
 \end{array}
\label{doublettripletsolutions}
\end{equation}
Since we know that in any realistic model, below the E.W. scale, the Higgs doublets spontaneously break the $SU(2) \times U(1)$ symmetry by acquiring vevs, we can look for SUSY solutions that allow for this mechanism to happen. If only the Higgs triplets are stabilized at the origin, there is only one solution that is:
\begin{equation}
 \begin{array}{ccc}
  \tilde{Y}_0=\frac{2 m_2}{3 \overline{\lambda}_{x}} diag(\{2,2,2,-3,-3\})&\tilde{\Phi} = \frac{2 m_2 }{\overline{\lambda}_{x}}\\
 \end{array}
\end{equation}
Along this direction the Higgs triplets are very heavy while the doublets remain massless (at tree-level):
\begin{equation}
  M=\frac{2 m_2}{3 \overline{\lambda}_{x}} diag(\{5,5,5,0,0\})
\end{equation}
We will now add to this extended version of the MSSM the model we presented in the second section, and assume that the solution which keeps the Higgs doublets light and the triplets heavy is the correct minimum of the potential.

The adjoints of $SU(5)$ will decompose to $(8,0)+(0,3)+(3,2)+(\overline{3},2)+$singlet. There is no mixing between the adjoints of the two sectors and they all get masses close to the GUT scale. The bifundamentals of the two sectors do mix: one chiral pair is eaten by the gauginos and becomes heavy, the two other pairs are ``light'' and get masses of the order of $\frac{F}{M}$ due to a see-saw structure of the mass matrix.

If we take the parameters to be:
\begin{table}[!h]
\begin{center}
 \begin{tabular}{|c|c|c|c|c|c|c|c|}
 \hline
$\mu$ & $\lambda$ & $\lambda_2$     &  $\frac{m}{\mu}$     & $\lambda_{x}$   & $\overline{\lambda}_{x}$ & $\frac{m_2}{\mu}$    & $g_{GUT}$\\[2mm]
 \hline
  \hline
$ 1$  &    $0.1$  &     $0.5$       &     $0.2887$         &   $0.0167$     &         $0.0025$          &       $0.05$         &    $1$\\
\hline
 \end{tabular}
\end{center}
\caption{Sample parameter point for the model with doublet-triplet slitting.}
\end{table}

The masses of the fermionic fields are (in units of $M_{GUT}$\footnote{$M_{GUT}$ is taken to be the mass of the Higgsed vector bosons}):

\begin{table}[!h]
\begin{center}
\begin{tabular}{|c||c|}
\hline
Adjoints $SU(3)$ & $(0.028,0.011,0.006)$\\
\hline
Adjoints $SU(2)$ & $(0.028,0.011,0.006)$\\
\hline
Higgs triplets  & $0.017$\\
\hline
Bifundamentals  & $(1,1,1.63 \times 10^{-9},6.03\times 10^{-10})$ \\
\hline
 \end{tabular}
\end{center}
\end{table}

So that if the GUT scale is $10^{16}GeV$, the two ``light'' bifundamentals would be around $10^7GeV$. We note that both the Higgs doublets and triplets have SUSY spectra, i.e. they do not couple to any F-terms at tree-level, as the vev of the F-term of $\tilde{\Phi}$ is 0. The Higgs sector is different from the quark and leptonic sectors since they can know about SUSY breaking indirectly through loops with $\tilde{\Phi}$ bifundamentals\footnote{So, even though the MSSM fields only know about SUSY breaking effects radiatively, strictly speaking this scenario is not pure gauge mediation.}.

Ignoring this effect which should be small if $\lambda_{x}$ and $\overline{\lambda}_{x}$ are small, we can give an order of magnitude estimate for the contribution coming from gauge mediation. One needs to take into account the one loop contribution from gauge messengers and the two loop contributions from both gauge and normal messengers. The one loop $F/M$ effect (given in (\ref{squarkmasses})) is $(F/M)^{(1)}=3,91\times 10^{-10}$.

The two loop $F/M$ effects are given by:
\begin{equation}
\begin{array}{c}
\frac{F}{M}_{g.m.}=1.12 \times 10^{-12}M_{GUT}\\
\frac{F}{M}_{n.m.}=4,02 \times 10^{-9}M_{GUT}\\
\end{array}
\end{equation}
So that the order of magnitude for the soft terms (at the messengers scale):
\begin{equation}
\begin{array}{c}
 m_{Q} \sim 10^{-11} M_{GUT}\\
 m_{L} \sim 10^{-11} M_{GUT}\\
m_{\lambda} \sim 4 \times 10^{-14}M_{GUT} \\
\end{array}
\end{equation}
Where there are nearly three orders of magnitude between gaugino and scalar masses. However, the values we present are computed at the messenger scale and R.G. effects are important and should be taken into account, and R.G. effects may be sufficient to solve this problem.

Even though this model allows us to use a solution to the doublet-triplet problem to get non-tachyonic scalar masses, and is very simple, it should be improved in order to become more realistic. A mechanism that makes the bifundamentals heavier would help reducing the tuning required to get unification:

\begin{figure}[h!]
\begin{center}
\includegraphics[scale=0.9]{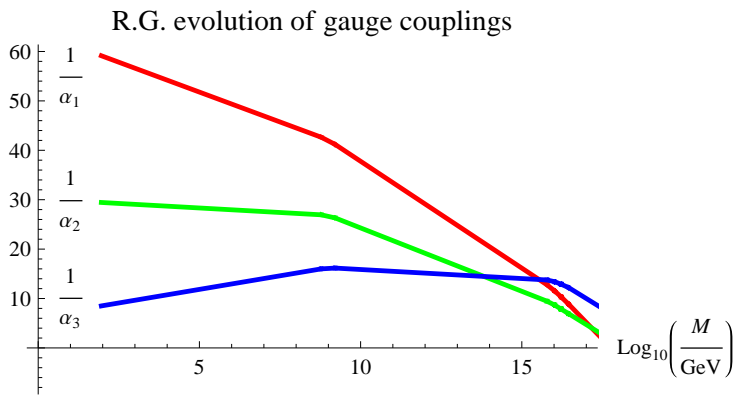}
\caption{One loop R.G. equations for the Gauge Couplings.}
\label{RGexample2}
\end{center}
\end{figure}

Where this plot is the R.G. evolution of the gauge couplings for the sample point we just presented.

\section{Conclusions}

In this paper we presented a model of F-term SUSY breaking with two $SU(5)$ adjoint chiral messenger fields and a singlet. Coupling these adjoint fields to the SUSY breaking sector broke the degeneracy between vacua with different symmetry breaking patterns. This gave us a natural mechanism that could explain why the $SU(5)$ GUT group is broken to the MSSM gauge group. In the particular model we presented this happened when one of the Yukawa couplings was $<1$ and positive.

In it's simplest form the model was not viable as quarks and sleptons were tachyonic. We showed that in a large class of models that have gauge messengers, this is associated with the need to stabilize the sgoldstino vev far away from the origin, and is independent of the values of the parameters.

To solve this problem we proposed two scenarios:

 SUSY is broken independently in two sectors, and gauge messengers exist in only one of them. There are two sgoldstinos that acquire different vevs. By choosing the ratio between these vevs it is possible to enhance the contributions from normal messengers and make both squark and sleptons non-tachyonic. We showed a concrete example where this scenario is realized, and that indeed squarks and sleptons can be non-tachyonic. Gaugino masses (and trilinear couplings) are generated at leading order in $F/M$ at one loop (because of gauge messengers), and up to R.G. effects are lighter than scalars (up to one order of magnitude).

In the second scenario there are also two sectors,but SUSY is only broken in one of them. The SUSY breaking sector should have both gauge and normal messengers, while the sector where SUSY is not broken should have a field whose vev breaks the GUT symmetry. If this vev is larger than the sgoldstino vev, the contribution from normal messengers can be enhanced and squarks and sleptons can be non-tachyonic.

One natural realization of this scenario is the sliding singlet solution to the doublet-triplet problem (or other solutions to the doublet-triplet problem) together with the first model we presented, as the hidden sector. We showed that with this extension of the MSSM, there exists a region of parameter space and field vevs for which both squarks and sleptons are non-tachyonic. This is not a complete model, however, and some of its problems were identified.

We also did not solve the problems with the Witten hierarchy idea\cite{Banks:1982mg}, but instead argued that it should be the SUSY breaking scale that is much lower than the GUT scale. This could be because of some approximate symmetry in the high energy theory: if unbroken, this symmetry would make the "natural" choice of parameters in the low energy model to be such that, despite the rank condition, SUSY is not broken.

\section*{Acknowledgements}

I would like to thank V. Khoze for useful discussions, and also e-mail exchanges with M. Sudano and K. Intriligator. This work was done with the support of a FTC studentship.

\newpage
\vspace{10mm}
\Large{\bf{Appendix:}}
\normalsize
\appendix

\section{Minimization of the potential:}
The model is:
\begin{equation}
  W=-\mu^2 \Phi+\lambda \Phi Tr(Y_0 Y_0)+mTr(Y_2 Y_0) +\overline{\lambda}Tr(Y_2 Y_0 Y_0)
\end{equation}
We shall describe the possible vevs that the Y fields can have using the generators of $SU(5)$. The generators of the Cartan subalgebra can be written as:
\begin{align}
T^{21}&=diag({1/2,-1/2,0,0,0})\nonumber\\
T^{22}&=diag({1/(2\sqrt{3}),1/(2\sqrt{3}),-1/(\sqrt{3}),0,0})\nonumber\\
T^{23}&=diag({1/(2\sqrt{6}),1/(2\sqrt{6}),1/(2\sqrt{6}),-3/(2\sqrt{6}),0})\\ T^{24}&=diag({1/(2\sqrt{10}),1/(2\sqrt{10})1/(2\sqrt{10}),1/(2\sqrt{10}),-2/\sqrt{10}})\nonumber\\
\end{align}
 The other generators can simply be written with the help of the $SU(2)$ generators.

Then the Y-fields can be written as
\begin{equation}
 Y_j=\sqrt{2} y_j^{k}T^{k}
\end{equation}
Where we take the $\sqrt{2}$ factor to so as to canonically normalize fields:
\begin{equation}
 K\supset Tr((Y_{j}^{\dagger}Y_{j})=2(y_j^{k})^{\dagger}y_j^{l}Tr(T^{k}T^{l})=(y_j^{k})^{\dagger}y_j^{l}
\end{equation}
We can now use the gauge degrees of freedom to align the $Y_0$ vevs along the directions spanned by the Cartan subalgebra (i.e. along the diagonal). We shall use letters from the middle of the alphabet (usually k) to mean that the field corresponds to a direction of the Cartan subalgebra of $SU(5)$, and a letter from the end of the alphabet to mean that the field is not along a direction spanned by the Cartan subalgebra (usually r).

We now analyze the F-term equations, we shall start by looking at the F-terms that are not along the Cartan subalgebra directions. By virtue of the gauge choice, the F-term equations for the $y_2^{r}$ directions vanish (not for the Cartan subalgebra directions), and the F-term equations for the $y_0^{r}$ fields also take a very simple form:
\begin{align}
W_{y_2^{r}}&=0\\
W_{y_0^{r}}&=f_{r}(\{y_0^{k}\})y_2^{r}
\end{align}
Where $f_1(y_0^{21},y_0^{22},y_0^{23},y_0^{24})=m+\frac{1}{15}\overline{\lambda}(5\sqrt{6}y_0^{22}+5\sqrt{3}y_0^{23}+3\sqrt{5}y_0^{24})$, and different f's have different expressions. So, the solution to these equations is given by $y_2^{r}=0$. The reason why we can do this for these fields is that they do not appear in any other F-term equations, so $V_F$ will have a quadratic term in these $y_2^{r}$ with a positive semi-definite coefficient given by $|f_{r}|^2$.

So, by choosing the $Y_0$ to be diagonal,one gets that due to the F-term equations $Y_2$ is also diagonal.

We will now subdivide the problem of minimizing $V_F=\sum W_i W^{i}$ into three cases: The symmetry breaking pattern is $SU(5)\rightarrow SU(4)\times  U(1)^2$, $SU(5)\rightarrow SU(3)\times  SU(2) \times  U(1)$ and generically when:
\begin{equation}
\begin{array}{cc}
\lambda\ll &1\\
\frac{\lambda}{\overline{\lambda}^2}<&1
\end{array}
\end{equation}
We start by analyzing the case: $SU(5)\rightarrow SU(4)\times  U(1)^2$.
This particular case is equivalent to choosing the vevs to lie along the direction spanned by $T^{24}$, so there are only three complex variables in the problem: $y_0^{24}$ and $y_2^{24}$ and $\Phi$. We use R-symmetry to choose $\Phi$ to be real. so we can write $y_0^{24}=y_{0r}+i\ y_{0i}$, $y_2^{24}=y_{2r}+ i\ y_{2i}$ and $\Phi$ is a real variable now.

Then the $V_F$ potential is given by:
\begin{eqnarray}
 V_F=&\mu^4+m^2(y_{0r}^2+y_{0i}^2+y_{2r}^2+y_{2i}^2)-\frac{3\overline{\lambda}}{\sqrt{5}}m\  y_{0r}(y_{0r}^2+y_{0i}^2+2(y_{2r}^2+y_{2i}^2))+\nonumber\\
&+\overline{\lambda}^2\frac{9}{20}(y_{0r}^2+y_{0i}^2)(y_{0r}^2+y_{0i}^2+4(y_{2r}^2+y_{2i}^2))+\nonumber\\
&+\lambda(4 m \Phi(y_{2r}\ y_{0r}+y1i\ y_{2i})-2 \mu^2 (y_{0r}^2-y_{0i}^2)-\frac{12}{\sqrt{5}} \overline{\lambda} \Phi y_{2r}(y_{0r}^2+y_{2i}^2))\nonumber\\
&+\lambda^2(y_{0r}^2+y_{0i}^2+4 \Phi^2)(y_{0r}^2+y_{0i}^2)
\end{eqnarray}
One then has to find the extremes of the potential by solving the system of equations corresponding to setting the gradient of the potential to 0 and check wether the solutions one find are local minima or maxima.
One gets the following set of solutions:
\begin{align}
solution_1=\{&y_{0r}=0,y_{0i}=0,y_{2r}=0,y_{2i}=0\}\\
 solution_{2,3}=\{&y_{0r}=\frac{\sqrt{5}(9m \overline{\lambda} \pm\sqrt{m^2(9\overline{\lambda}^2-160\lambda^2)+16\lambda \mu^2(20\lambda^2+9\overline{\lambda})})}{40\lambda^2+18\overline{\lambda}^2},y_{0i}=0,y_{2i}=0,\nonumber\\
&\Phi=y_{2r}(\frac{-3m^2\overline{\lambda}m^2+24\lambda\overline{\lambda}\mu^2\pm m\sqrt{m^2(9\overline{\lambda}^2-160\lambda^2)+16\lambda \mu^2(20\lambda^2+9\overline{\lambda}}}{8\sqrt{5}\lambda(2\lambda \mu^2-m^2)}\}\\
solution_{4,5}=\{&y_{0r}=\frac{3\sqrt{5} m\overline{\lambda}(m^2+2\lambda \mu^2}{80 \lambda^3 \mu^2+9 \overline{\lambda}^2(m^2+4\lambda \mu^2)},\nonumber\\
&y_{0i}= \pm\sqrt{5} \sqrt{-\frac{(m^2+2\lambda \mu^2)(9 \overline{\lambda}^2 m^4+90\lambda \overline{\lambda}m^2 \mu^2+32 \lambda^2(20\lambda^2+9\overline{\lambda})\mu^4)}{(80\lambda^3 \mu^2+9\overline{\lambda}^2(m^2+4\lambda \mu^2))^2}},\nonumber\\
&y_{2i}=\frac{5m y_{0i}}{5my_{0r}-3\sqrt{5}\overline{\lambda}(y_{0i}^2+y_{0r}^2)}y_{2r},\nonumber\\
&\Phi=-\frac{5m^2-6\sqrt{5} \overline{\lambda} m y_{0r}+9\overline{\lambda}^2(y_{0r}^2+y_{0i}^2)}{2\lambda(5my_{0r}-3\sqrt{5}\overline{\lambda}(y_{0r}^2+y_{0i}^2))}y_{2r}\}
\vspace{-0.7cm}
\end{align}
Where we note that solutions 4 and 5 do not exist for small $\lambda$ (and if $\lambda<0$ these solutions are even more complicated).

The value of $V_F$ for solutions 1,2 and 3 is:
\begin{align}
\vspace{-0.7cm}
 V_F^{(1)}=&\mu^4\\
V_F^{(2,3)}=&\frac{1}{8(20\lambda^2+\overline{\lambda}^2)^3}(5m^4(81\overline{\lambda}^4+3600\lambda^2\overline{\lambda}^2-3200\lambda^4)+200m^2 \lambda(320\lambda^4-36\lambda^2\overline{\lambda}^2-81\overline{\lambda}^4)\mu^2+\nonumber\\
&+72(20\lambda^2 \overline{\lambda}+9\overline{\lambda}^3)^2\mu^4 \pm (15 m \overline{\lambda}((9\overline{\lambda}^2-160\lambda^2)m^2+16\lambda(20\lambda^2+9\overline{\lambda}^2)\mu^2)^{3/2}
\vspace{-0.7cm}
\end{align}
Where we note that solutions 2 and 3 are not equivalent and have different values for the potential at the minimum. One can now expand these solutions to linear order in $\lambda$ and get the results mentioned in section 1.

The analysis of the case $SU(5)\rightarrow SU(3)\times  SU(2) \times  U(1)$ is very similar. We will just outline the main differences.
The vevs are now chosen to be along a combination of the directions given by $T^{23}$ and $T^{24}$: $y_0^{23}=\sqrt{5/3}y_0^{24}$, $y_2^{23}=\sqrt{5/3}y_2^{24}$. Again there are only 3 complex valued variables, and we can use R-symmetry to make one of them be real:
$y_0^{24}=y_{0r}+i\ y_{0i}$, $y_2^{24}=y_{2r}+ i\ y_{2i}$ and $\Phi$ is real.

The extremes of the potential are:
\begin{align}
 solution_1=\{&y_{0r}=0,y_{0i}=0,y_{2r}=0,y_{2i}=0\}\\
solution_{2,3}=\{&y_{0r}=\frac{3\sqrt{5}}{8(30\lambda^2+\overline{\lambda}^2)}(3m\overline{\lambda}\pm\sqrt{m^2(\overline{\lambda}^2-240\lambda^2)+16\lambda \mu^2(\overline{\lambda}+30\lambda^2)}),y_{0i}=0,\nonumber\\
&y_{2i}=0,\Phi=-\frac{45m^2-24\sqrt{5} \overline{\lambda} m y1r+16 \overline{\lambda}(y_{0r}^2+y_{0i}^2)}{6\lambda(15m y_{0r}-4\sqrt{5}\overline{\lambda}(y_{0r}^2+y_{0i}^2)}y_{2r}\}\\
solution_{4,5}=\{&y_{0r}=\frac{3\sqrt{5}m \overline{\lambda}(m^2+2\lambda \mu^2)}{4(\overline{\lambda}^2 m^2+4 \lambda \mu^2 (30 \lambda^2+\overline{\lambda}^2))},\nonumber\\
&y_{0i}=\pm\frac{3\sqrt{5}}{4}\sqrt{-\frac{(m^2+2\lambda \mu^2)(\overline{\lambda}^4 m^4+10 \lambda \overline{\lambda}^2 m^2 \mu^2+32 \lambda^2 \mu^4 (30\lambda^2+\overline{\lambda}^2))}{(m^2 \overline{\lambda}^2+4 \lambda \mu^2 (30 \lambda^2+\overline{\lambda}^2))^2}},\nonumber\\
&y_{2i}=\frac{15 m y_{0r}}{15 m y_{0r}-4\sqrt{5}\overline{\lambda}(y_{0r}^2+y_{0i}^2)}y_{2r},\nonumber\\
&\Phi=-\frac{45m^2-24\sqrt{5} \overline{\lambda} m y1r+16 \overline{\lambda}(y_{0r}^2+y_{0i}^2)}{6\lambda(15m y_{0r}-4\sqrt{5}\overline{\lambda}(y_{0r}^2+y_{0i}^2)}y_{2r}\}
\vspace{-0.5cm}
\end{align}
As in the previous case these two last solutions only exist when $\lambda$ is not small.
The potential at the other three solutions is given by:
\begin{align}
 V_F^{(1)}=&\mu^4\\
V_F^{(2,3)}=&\frac{1}{16(20\lambda^2+\overline{\lambda}^2)^3}(15m^4(\overline{\lambda}^4+600\lambda^2\overline{\lambda}^2-7200\lambda^4)+600m^2 \lambda(720\lambda^4-6\lambda^2\overline{\lambda}^2-\overline{\lambda}^4)\mu^2+\nonumber\\
&16(30\lambda^2 \overline{\lambda}+\overline{\lambda}^3)^2\mu^4 \pm (15 m \overline{\lambda}((\overline{\lambda}^2-240\lambda^2)m^2+16\lambda(30\lambda^2+\overline{\lambda}^2)\mu^2)^{3/2}
\vspace{-0.4cm}
\end{align}
And if we expand these values we get the results quoted in section 1.

We now turn to the general case where no preferred symmetry breaking pattern is chosen. The first thing is to rescale all the Y-fields by $\frac{m}{\overline{\lambda}}$, as this simplifies the potential (if $\lambda=0$ this makes the function to minimize independent of any parameter in the model).

Then we not that the fields $y_2^{k}$, due to their R-charge, can only appear in the $F_{y_0^{k}}$ terms. These are n-equations in n-variables (the $y_2^{k}$'s) and can be solved. This obviously does not hurt the global minimization of the potential. The solutions are quite long and not particularly deep, so we shall not present them here. The bottom line is that one has only to consider the F-terms of the fields with R-charge equal to 2 in the minimization of the potential.

The potential can then be written as:
\begin{equation}
\footnotesize
\!\!\!\!
\begin{array}{c}
 V_F=\frac{m^4}{\overline{\lambda}^2} v(y_0^{21},y_0^{22},y_0^{23},y_0^{24})-2\lambda \frac{\mu^2 m^2}{\overline{\lambda}^2}Re((y_0^{21})^2+(y_0^{22})^2+(y_0^{23})^2+(y_0^{24})^2)+\mu^4+O((\frac{\lambda}{\overline{\lambda}^2})^2)
 \end{array}
 \normalsize
\end{equation}
Where $v(x,y,z,w)$ is a complicated positive semi-definite function whose minimum is 0. The reason this must be so is that when $\lambda=0$ the messengers decouple from the SUSY breaking sector, so their F-terms vanish in the vacuum.
We will now use perturbation theory to minimize this potential where we will take $\frac{\lambda}{\overline{\lambda}^2}<1$ as the small parameter.
The minima of the function v can be found exactly since this amount to solving $F_{y_2^{k}}=0$ when $\lambda=0$, i.e.
\begin{align}
&y_0^{21}(15+5\sqrt{6}y_0^{22}+5\sqrt{3}y_0^{23}+3\sqrt{5}y_0^{24})=0\\
&5\sqrt{6}(y_0^{21})^2+y_0^{22}(30-5\sqrt{6}y_0^{22}+10\sqrt{3}y_0^{23}+6\sqrt{5}y_0^{24})=0\\
&5\sqrt{3}(y_0^{21})^2+5\sqrt{3}(y_0^{22})^2+2y_0^{23}(15-5\sqrt{3}y_0^{23}+3\sqrt{5}y_0^{24})=0\\
&\sqrt{5}(y_0^{21})^2+\sqrt{5}(y_0^{22})^2++\sqrt{5}(y_0^{23})^2+y_0^{24}(10-3\sqrt{5}y_0^{24})=0
\end{align}
This system of equations has $2^4=16$ solutions out of which only three are independent:
\begin{align}
&Y_0=diag(\{0,0,0,0,0\})\\
&Y_0=\frac{m}{\overline{\lambda}} diag(\{2,2,2,-3,-3\})\\
&Y_0=\frac{m}{3\overline{\lambda}} diag(\{1,1,1,1,-4\})
\end{align}
The rest being permutations of these solutions(there are $1+\binom{5}{2}+5=16$ of these).

One can then use perturbation theory and expand around the solutions we found to linear order in $\frac{\lambda}{\overline{\lambda}^2}$ to get approximate solutions to the potential. Since these approximations will respect one of the symmetry breaking patterns we have already studied we shall not repeat this operation here.

\vspace{-0.1cm}
\section{The messenger mass matrices}

Let us now assume that $\lambda>0$, and focus on the case when the GUT group is broken down to the MSSM gauge group.
Due to the Higgsing of the GUT group the adjoints of SU(5) decompose under the unbroken $SU(3)\times  SU(2) \times  U(1)$ as $(8,1)\times (1,3)\times (3,2)\times (\overline{3},2)\times (1,1)$, i.e. one adjoint of $SU(3)$, one adjoint of $SU(2)$,a vector-like pair of bifundamentals and a singlet.
The field content of the model then becomes:
{\renewcommand{\arraystretch}{1.8}
\renewcommand{\tabcolsep}{0.4cm}
\begin{table}[h]
\begin{center}
\begin{tabular}{|c | c | c |}
\hline
$GUT$  &  $SU(3)\times  SU(2) \times  U(1)$     &    $Representation$\\
\hline
\hline
$V$    &      $\left(\begin{array}{cc}
		V_{SU(3)}  &  \psi  \\
		\tilde{\psi} & V_{SU(2)} \\
		\end{array}\right)+\rho1$   &    $ \left(\begin{array}{cc}
						(8,1)  &  (3,2)  \\
						(\overline{3},2) & (0,3) \\
						\end{array}\right)+singlet$\\
\hline

$Y_2$    &      $\left(\begin{array}{cc}
		Y_{2,adj3}  &  \tau  \\
		\tilde{\tau} & Y_{2,adj2} \\
		\end{array}\right)+\rho2$   &    $ \left(\begin{array}{cc}
						(8,1)  &  (3,2)  \\
						(\overline{3},2) & (0,3) \\
						\end{array}\right)+singlet$\\
\hline

$Y_0$    &      $\left(\begin{array}{cc}
		Y_{0,adj3}  &  \chi  \\
		\tilde{\chi} & Y_{0,adj2} \\
		\end{array}\right)+\rho3$    &    $ \left(\begin{array}{cc}
						(8,1)  &  (3,2)  \\
						(\overline{3},2) & (0,3) \\
						\end{array}\right)+singlet$\\
\hline
$\Phi$    &      $\Phi$                       &    $singlet$\\
\hline
\end{tabular}
\end{center}
\label{fielddefinitions}
\caption{Field Decomposition and notation.}
\end{table}}

Let us now get the fermionic and bosonic masses for the particles in the model. We shall focus in the region of parameter space where $\lambda < 1, \overline{\lambda} y > m$.

 If we focus first in the fermionic mass matrix for the adjoint fields of the unbroken $SU(3)$, (since we saw that all vevs are real values, we shall drop complex conjugation symbols to simplify the formulas) we get that the fermionic mass matrix for these fields is given by:
\begin{align}
 ((Y^{a}_{2,adj3})^{\dagger}, (Y^{a}_{0,adj3})^{\dagger}) (M_f^{2,(adj)}) \left(\begin{array}{c}
                        Y^{a}_{2,adj3} \\
			 Y^{a}_{0,adj3}
                        \end{array} \right)
\end{align}
And $Y_{2,adj3}=\sum Y^{a}_{2,adj3}\lambda^{a}$, where $\lambda^{a}$ is a basis of the $SU(3)$ algebra.
Where $M_f^{2,(adj)}$ is, at leading order in $\lambda$ given by:
\begin{equation}
  M_f^{2,(adj)}=\left( \begin{array}{cc}
                       25 m^2+\frac{20}{9} \overline{\lambda} y+ \frac{16}{9} \lambda\mu^2 (\frac{\overline{\lambda}^2 y^2}{m^2}+45) &\frac{10\sqrt{5}}{3} \overline{\lambda} y m+\frac{20\sqrt{5}}{3}\lambda \mu^2 \frac{\overline{\lambda} y}{m3} \\
\frac{10\sqrt{5}}{3} \overline{\lambda} y m+\frac{20\sqrt{5}}{3}\lambda \mu^2 \frac{\overline{\lambda} y}{m3} &  25 m^2+80 \lambda \mu^2 \\
                     \end{array}
\right)
\end{equation}
To linear order in $\lambda$ and second order in $\frac{1}{\overline{\lambda} y}$ we get:
\begin{align}
&m_{H}^2=50m^2+160 \lambda \mu^2+\frac{4}{9}(5 m^2+4 \lambda \mu^2) \frac{\overline{\lambda} ^2 y^2}{m^2}-\frac{225m^2}{4} \frac{5 m^2 +28 \lambda \mu^2}{\overline{\lambda}^2 y^2}\\
&m_{l}^2=\frac{225}{4} m^2 \frac{5 m^2+28 \lambda \mu^2}{\overline{\lambda}^2 y^2}
\end{align}
The scalar mass matrix can be written as:
\begin{align}
 ((Y^{a}_{2,adj3})^{\dagger},\ (Y^{a}_{0,adj3})^{\dagger},\ (Y^{a}_{2,adj3})^{T},\ (Y^{a}_{0,adj3})^{T}) (M_b^{2,(adj)}) \left(\begin{array}{c}
                        Y^{a}_{2,adj3} \\
			Y^{a}_{0,adj3}\\
 			(Y^{a}_{2,adj3})^{*}\\
 			(Y^{a}_{0,adj3})^{*}
                        \end{array} \right)
\end{align}
Or replacing the vevs:
\scriptsize
\begin{equation*}
\!\!\!\!\!\!\!\!  M_b^{2,(adj)}={\left(\begin{array}{cccc}
                       25 m^2+80 \lambda \mu^2 +\frac{4}{9} \frac{\overline{\lambda}^2 y^2(5m^2+4 \lambda \mu^2)}{m^2} & \frac{10\sqrt{5}}{3} \frac{\overline{\lambda} y(m^2+2 \lambda \mu^2)}{ m} &-10\lambda \mu^2  &0\\
\frac{10\sqrt{5}}{3} \frac{\overline{\lambda} y(m^2+2 \lambda \mu^2)}{m} & 25 m^2+80 \lambda \mu^2 & 0 & 0 \\
-10\lambda \mu^2 & 0 &  25 m^2+80 \lambda \mu^2 +\frac{4}{9} \frac{\overline{\lambda}^2 y^2(5m^2+4 \lambda \mu^2)}{m^2} & \frac{10\sqrt{5}}{3} \frac{\overline{\lambda} y(m^2+2 \lambda \mu^2 )}{m} \\
0 & 0 &  \frac{10\sqrt{5}}{3} \frac{\overline{\lambda} y(m^2+2 \lambda \mu^2 )}{ m}& 25 m^2+80 \lambda \mu^2  \\
\end{array} \right)}
\end{equation*}
\normalsize
And the eigenvalues are given by:
\begin{align}
&\footnotesize
\begin{array}{c}
m_{H,\pm}^2=50m^2+160 \lambda \mu^2+\frac{4}{9}(5 m^2+4 \lambda \mu^2) \frac{\overline{\lambda}^2 y^2}{m^2}-\frac{225m^2}{4} \frac{5 m^2+28 \lambda \mu^2}{\overline{\lambda}^2 y^2} \pm \frac{5 \lambda \mu^2(4 \overline{\lambda}^2 y^2-45 m^2}{2y^2\overline{\lambda}^2 })
\end{array}
\normalsize
\\
&
\footnotesize
\begin{array}{c}
m_{l,\pm}^2=\frac{225m^2}{4} \frac{5 m^2 +28 \lambda \mu^2}{\overline{\lambda}^2 y^2}\pm \frac{225}{2}\frac{\lambda m^2 \mu^2}{\overline{\lambda}^2  y^2}
\end{array}
\normalsize
\end{align}
\normalsize
For the adjoints of $SU(2)$ the mass matrix is very similar, the fermion masses are:
\begin{align}
&
m_{H}^2=50m^2+240 \lambda \mu^2+\frac{4}{9}(5 m^2-4 \lambda \mu^2) \frac{\overline{\lambda}^2 y^2}{m^2 }-\frac{225m^2}{4} \frac{5 m^2 +52 \lambda \mu^2}{\overline{\lambda}^2 y^2}\\
&m_{l}^2=\frac{225m^2}{4} \frac{5 m^2 +52 \lambda \mu^2}{\overline{\lambda}^2 y^2}
\end{align}
And the scalar masses are:
\begin{align}
&\footnotesize
\begin{array}{c}m_{H}^2=50m^2+240 \lambda \mu^2+\frac{4}{9}(5 m^2-4 \lambda \mu^2) \frac{\overline{\lambda}^2y^2}{m^2 }-\frac{225m^2}{4} \frac{5 m^2 +52 \lambda \mu^2}{\overline{\lambda}^2 y^2} \pm \frac{5 \lambda \mu^2(4\overline{\lambda}^2 y^2 -45 m^2)}{\overline{\lambda}^2 y^2}
\end{array}\\
&\footnotesize
\begin{array}{c}
m_{l}^2=\frac{225m^2}{4} \frac{5 m^2 +52 \lambda \mu^2}{\overline{\lambda}^2 y^2}\pm \frac{225}{2} \frac{\lambda  \mu^2 m^2}{\overline{\lambda}^2 y^2}
\end{array}
\end{align}
So that for large values of the pseudomodulos vev one of the fields gets heavy while the other gets light.
This is a simple consequence of R-symmetry together with the fact that these adjoints do not enter in the Higgs mechanism (i.e. they only get masses through the superpotential). Since the superpotential is R-symmetric, one can generically show that the $det(M)$ does not depend of the vev of the scalar partner of the goldstino y (i.e our flat direction). Since $Tr(M)$ does depend on y, it has to be that for large values of y one of the eigenvalues has to go with $m^{1-r} y^{r}$ while the other goes as $(m^{1+r} y^{-r}$, for some value of r. This gives us a sort of see-saw mechanism where as y increases one of the field becomes lighter and the other becomes heavier.

The complete fermionic mass matrices (we take all vevs to be real):

Adjoint of $SU(3)$
\begin{equation}
 (Y^{a}_{2,adj3}, Y^{a}_{0,adj3}) M_f^{(adj)} \left(\begin{array}{c}
                        Y^{a}_{2,adj3} \\
			 Y^{a}_{0,adj3}
                        \end{array} \right)
\end{equation}
Where:
\begin{equation*}
\tiny
 M_f^{(adj)}=\left(
\begin{array}{cc}
 \frac{1}{45}(45m^2+48\sqrt{5}m \overline{\lambda} y_0^{24}+180 \lambda^2 \Phi^2+96\sqrt{5} \lambda \overline{\lambda} \Phi y_2^{24}+64 \overline{\lambda}^2((y_0^{(24)})^2+(y_2^{(24)})^2) & \frac{2}{225}(15 m+8\sqrt{5} \overline{\lambda} y_0^{(24)})(15\lambda\Phi+4\sqrt{5} \overline{\lambda} y_2^{(24)}) \\
 \frac{2}{225}(15 m+8\sqrt{5} \overline{\lambda} y_0^{(24)})(15\lambda\Phi+4\sqrt{5} \overline{\lambda} y_2^{(24)}) & \frac{1}{225}(15m+8\sqrt{5}\overline{\lambda} y_0^{(24)})^2
\end{array}
\right)
\end{equation*}
\normalsize
(The adjoints $V_{SU(3)}$ of the unbroken $SU(3)$ are obviously massless)

The eigenvalues can be computed and are:
\begin{align}
\!\!\!\!\!\!\!\!\small\begin{array}{c}
mass_{1,2}=\end{array}& \begin{array}{c}\frac{1}{45}((8\overline{\lambda} y_0^{(24)}+3\sqrt{5} m)^2+(2\sqrt{2} \overline{\lambda} y_2^{(24)}-\sqrt{10} \lambda \Phi)^2 \end{array}\nonumber\\
&\begin{array}{c}\pm Abs(2\sqrt(2)\overline{\lambda}y_2^{(24)}-\sqrt{10} \lambda \Phi)\sqrt{2(4 \overline{\lambda} y_0^{(24)}-\sqrt{5} m)^2+(2\sqrt(2) \overline{\lambda}-\sqrt{10} \lambda \Phi)^2}\end{array}
\normalsize
\end{align}
Adjoint of $SU(2)$
\begin{equation}
 (Y^{a}_{2,adj2}, Y^{a}_{0,adj2}) M_f^{(adj2)} \left(\begin{array}{c}
                        Y^{a}_{2,adj2} \\
			 Y^{a}_{0,adj2}
                        \end{array} \right)
\end{equation}
Where:
\begin{equation*}
\tiny
 M_f^{(adj2)}=\left(
\begin{array}{cc}
 m^2-\frac{8\overline{\lambda} y_0^{(24)}m}{\sqrt{5}}+4\lambda^2 \Phi^2 +\frac{16}{5}\overline{\lambda}(\overline{\lambda}((y_0^{(24)})^2+(y_2^{(24)})^2)-\sqrt{5}\lambda \Phi y_2^{(24)}) & -\frac{2}{225}(5 m-4\sqrt{5} \overline{\lambda} y_0^{(24)})(-5\lambda\Phi+2\sqrt{5} \overline{\lambda} y_2^{(24)}) \\
 -\frac{2}{225}(5 m-4\sqrt{5} \overline{\lambda} y_0^{(24)})(-5\lambda\Phi+2\sqrt{5} \overline{\lambda} y_2^{(24)}) & \frac{1}{25}(5m-4\sqrt{5}\overline{\lambda} y_0^{(24)})^2\\
\end{array}
\right)
\end{equation*}
(The adjoints $V_{SU(2)}$ of the unbroken $SU(2)$ are obviously massless)

The eigenvalues can be computed and are:
\begin{align}
\!\!\!\!\!\!\!\!\small\begin{array}{c}
 mass_{1,2}=\end{array}&\begin{array}{c}\frac{1}{5}((4\overline{\lambda} y_0^{(24)}-\sqrt{5} m)^2+(2\sqrt{2} \overline{\lambda} y_2^{(24)}-\sqrt{10} \lambda \Phi)^2\end{array} \nonumber\\
&\begin{array}{c}\pm Abs(2\sqrt(2)\overline{\lambda}y_2^{(24)}-\sqrt{10} \lambda \Phi)\sqrt{2(4 \overline{\lambda} y_0^{(24)}-\sqrt{5} m)^2+(2\sqrt(2) \overline{\lambda}-\sqrt{10} \lambda \Phi)^2}\end{array}
\end{align}
if $y_2^{24}$ is sufficiently large, the splittings in the scalar mass matrices are small, so that $F/M$ can be computed as the mass splitting over the mass (i.e. this is exact up to $(F/M)^2$ corrections). By doing this one gets that:
\begin{equation}
 \frac{F}{M}=\frac{3\sqrt{5} \lambda \mu^2}{\overline{\lambda} y_2^{24}}
\end{equation}
for both gauge and non-gauge messengers (for gauge messengers one has to do this for the fermionic mass matrix, but the discussion goes through almost word by word).
\vspace{-0.2cm}
\section{Soft terms in models with two mass thresholds}
\label{RGtwothresholds}

Let us now compute the leading order contribution to squark masses assuming that there are two mass thresholds (and one SUSY breaking scale). One of the thresholds concerns normal (non-gauge messengers) and the other is with respect to normal gauge messengers. Here we will assume that gauge messengers are heavier than non-gauge messengers.

The computation of the running gauge function is done by solving the beta-function R.G. equations, one gets three solutions: one above the GUT scale, one between the GUT scale (where the gauge messengers are integrated out) and the non-gauge messengers, and one below the scale of the non-gauge messengers.
These are respectively:
\begin{align}
&\alpha^{-1}(\mu)=\alpha^{-1}(\Lambda_{U.V.})+\frac{b''}{4\pi}Log(\frac{\mu}{\Lambda_{U.V.}})\\
&\alpha^{-1}(\mu)=\alpha^{-1}(\Lambda_{U.V.})+\frac{b''}{4\pi}Log(\frac{(X,X)}{\Lambda_{U.V.}})+\frac{b'}{4\pi}Log(\frac{\mu^2}{(X,X)})\\
&\alpha^{-1}(\mu)=\alpha^{-1}(\Lambda_{U.V.})+\frac{b''}{4\pi}Log(\frac{\mu}{\Lambda_{U.V.}})+\frac{b'}{4\pi}Log(\frac{X_2^{\dagger}X_2}{(X,X)})+\frac{b}{4\pi}Log(\frac{\mu^2}{X_2^{\dagger}X2})
\end{align}
Where $b''$,$b'$,$b$ are the gauge function beta-function coefficients in the three different regimes. And the mass of the gauge messengers is $\sqrt{(X,X)}$, and the mass of the normal messengers is $X_2$.

We shall introduce now the following notation for the real gauge coupling and the squark wave-function renormalization functions and spurion-like fields: if they have a subscript s they are to be understood as the analytically continued functions into superspace, while if they do not have an s subscript, they are ordinary (scalar part) functions.

Upon continuation to superspace one gets:
\begin{equation}
\begin{array}{c}
 (X_s,X_s)=(x,x)+\theta^2(x,F)+\overline{\theta}^2(F,x)+\overline{\theta}^2 \theta^2 (F,F)\\
X_{2s}=x_2+\theta^2 F_2
\end{array}
\end{equation}
And $x_2$ is the mass scale of the non-gauge messengers and $F_2$ is the component of the goldstino that they couple to. For the gauge messengers X is the goldstino and F is the vev of the F-term associated.

We can now see how the expectation values of the auxiliary components of the real gauge coupling get vevs upon this analytic continuation into superspace.
Replacing the definitions of $(X_s,X_s)$ and $X_{2s}$ ``spurions'' into the equations of the real gauge couplings we get:
\begin{align}
\alpha_s^{-1}(M)=& \alpha^{-1}(M)+\frac{b''}{4\pi}Log(1+\theta^2 \frac{(x,F)}{(x,x)}+\overline{\theta}\frac{(F,x)}{x,x}+\theta^2\overline{\theta}^2 \frac{(F,F)}{(x,x)})\\
\alpha_s^{-1}(x_2)=& \alpha^{-1}(x_2)+\frac{b''-b'}{4\pi}Log(1+\theta^2 \frac{(x,F)}{(x,x)}+\overline{\theta}\frac{(F,x)}{(x,x)}+\theta^2\overline{\theta}^2 \frac{(F,F)}{(x,x)})\nonumber\\
&+\frac{b'}{4\pi}Log(1+\theta^2 \frac{F_2}{x_2^{\dagger}}+\overline{\theta}\frac{F^{\dagger}}{X}+\theta^2\overline{\theta}^2 \frac{|F|^2}{|x_2|^2})\\
\alpha_s^{-1}(\mu)= &\alpha^{-1}(\mu)+\frac{b''-b'}{4\pi}Log(1+\theta^2 \frac{(x,F)}{(x,x)}+\overline{\theta}\frac{(F,x)}{(x,x)}+\theta^2\overline{\theta}^2 \frac{(F,F)}{(x,x)})\nonumber\\
&+\frac{b'-b}{4\pi}Log(1+\theta^2 \frac{F_2}{x_2^{\dagger}}+\overline{\theta}\frac{F^{\dagger}}{X}+\theta^2\overline{\theta}^2 \frac{|F|^2}{|x_2|^2})+\frac{b}{4\pi}Log(1+\theta^2 \frac{F2}{x2^{\dagger}}+\overline{\theta}\frac{F^{\dagger}}{X}+\theta^2\overline{\theta}^2 \frac{|F|^2}{|x_2|^2})
\end{align}
One can now solve the R.G. equation for the wave function-renormalization $Z_Q$, and get that:
\begin{equation}
 -Log(Z(\mu))=2 \frac{c''}{b''}Log(\frac{\alpha(m_v)}{\alpha(\Lambda_{U.V.})})+2 \frac{c'}{b'}Log(\frac{\alpha(x_2)}{\alpha(m_v)})+2 \frac{c'}{b}Log(\frac{\alpha(\mu)}{\alpha(x_2)})
\end{equation}
Replacing the expression for the running gauge couplings in this expression and recalling that in a general model
\begin{equation}
 m^2_Q=-Log(Z(\mu))|_{\theta^2 \overline{\theta}^2}
\end{equation}
We get that:
\begin{align}
 m^2_Q=&\frac{\alpha(\mu)}{2\pi}(c'-c''-2c'\frac{b''}{b'}-\chi1(c' \frac{b''}{b'}+c'')+\chi2(b-b')(1-\frac{b''}{b'})c')\frac{(F,F)(x,x)-(F,x)(x,F)}{(x,x)^2}\nonumber\\
&+\frac{\alpha(\mu)^2}{8\pi^2}( (b-b')c'\frac{|F_2|^2}{|x_2|^2}+(b'c'+b''(c''-2c')+2c' b'\frac{b'}{b})\frac{(x,F)(F,x)}{(x,x)^2}+2\chi1b''(c'\frac{b''}{b'}+c'')\frac{(F,x)(x,F)}{(x,x)^2}\nonumber\\
&-2\chi2\frac{(1-\frac{b'}{b})c'(-F2^{\dagger}(x,x) b'+x2 (F,x)(b'-b''))(F2(x,x)b'+x2^{\dagger}(x,F)(b''-b'))}{|x2|^2(x,x)^2})+O(\chi1^2,\chi2^2)
\end{align}
Where $\chi1=\frac{\alpha(M)}{\alpha(\mu)}-1$,$\chi2=\frac{\alpha(x2)}{\alpha(\mu)}-1$. So that we see that if we set $\chi1$ and $\chi2$ to zero, we get a sum of the naive expectations for the masses.

\newpage
\vspace{-0.2cm}
\bibliographystyle{utphys}
\bibliography{bibliography}

\providecommand{\href}[2]{#2}\begingroup\raggedright\begin{thebibliography}{10}

\bibitem{Intriligator:2007py}
K.~A. Intriligator, N.~Seiberg, and D.~Shih, ``{Supersymmetry Breaking,
  R-Symmetry Breaking and Metastable Vacua},''
  \href{http://dx.doi.org/10.1088/1126-6708/2007/07/017}{{\em JHEP} {\bfseries
  07} (2007) 017},
\href{http://arxiv.org/abs/hep-th/0703281}{{\ttfamily arXiv:hep-th/0703281}}.

\bibitem{Meade:2008wd}
P.~Meade, N.~Seiberg, and D.~Shih, ``{General Gauge Mediation},''
  \href{http://dx.doi.org/10.1143/PTPS.177.143}{{\em Prog. Theor. Phys. Suppl.}
  {\bfseries 177} (2009) 143--158},
\href{http://arxiv.org/abs/0801.3278}{{\ttfamily arXiv:0801.3278 [hep-ph]}}.

\bibitem{Buican:2009vv}
M.~Buican and Z.~Komargodski, ``{Soft Terms from Broken Symmetries},''
  \href{http://dx.doi.org/10.1007/JHEP02(2010)005}{{\em JHEP} {\bfseries 02}
  (2010) 005},
\href{http://arxiv.org/abs/0909.4824}{{\ttfamily arXiv:0909.4824 [hep-ph]}}.

\bibitem{Abel:2009ve}
S.~Abel, M.~J. Dolan, J.~Jaeckel, and V.~V. Khoze, ``{Phenomenology of Pure
  General Gauge Mediation},''
  \href{http://dx.doi.org/10.1088/1126-6708/2009/12/001}{{\em JHEP} {\bfseries
  12} (2009) 001},
\href{http://arxiv.org/abs/0910.2674}{{\ttfamily arXiv:0910.2674 [hep-ph]}}.

\bibitem{Abel:2007nr}
S.~A. Abel, C.~Durnford, J.~Jaeckel, and V.~V. Khoze, ``{Patterns of Gauge
  Mediation in Metastable SUSY Breaking},''
  \href{http://dx.doi.org/10.1088/1126-6708/2008/02/074}{{\em JHEP} {\bfseries
  02} (2008) 074},
\href{http://arxiv.org/abs/0712.1812}{{\ttfamily arXiv:0712.1812 [hep-ph]}}.

\bibitem{Abel:2008gv}
S.~Abel, J.~Jaeckel, V.~V. Khoze, and L.~Matos, ``{On the Diversity of Gauge
  Mediation: Footprints of Dynamical SUSY Breaking},''
  \href{http://dx.doi.org/10.1088/1126-6708/2009/03/017}{{\em JHEP} {\bfseries
  03} (2009) 017},
\href{http://arxiv.org/abs/0812.3119}{{\ttfamily arXiv:0812.3119 [hep-ph]}}.

\bibitem{Carpenter:2008wi}
L.~M. Carpenter, M.~Dine, G.~Festuccia, and J.~D. Mason, ``{Implementing
  General Gauge Mediation},''
  \href{http://dx.doi.org/10.1103/PhysRevD.79.035002}{{\em Phys. Rev.}
  {\bfseries D79} (2009) 035002},
\href{http://arxiv.org/abs/0805.2944}{{\ttfamily arXiv:0805.2944 [hep-ph]}}.

\bibitem{Dine:2006xt}
M.~Dine and J.~Mason, ``{Gauge mediation in metastable vacua},''
  \href{http://dx.doi.org/10.1103/PhysRevD.77.016005}{{\em Phys. Rev.}
  {\bfseries D77} (2008) 016005},
\href{http://arxiv.org/abs/hep-ph/0611312}{{\ttfamily arXiv:hep-ph/0611312}}.

\bibitem{Dine:2007dz}
M.~Dine and J.~D. Mason, ``{Dynamical Supersymmetry Breaking and Low Energy
  Gauge Mediation},'' \href{http://dx.doi.org/10.1103/PhysRevD.78.055013}{{\em
  Phys. Rev.} {\bfseries D78} (2008) 055013},
\href{http://arxiv.org/abs/0712.1355}{{\ttfamily arXiv:0712.1355 [hep-ph]}}.

\bibitem{Intriligator:2010be}
K.~Intriligator and M.~Sudano, ``{General Gauge Mediation with Gauge
  Messengers},'' \href{http://dx.doi.org/10.1007/JHEP06(2010)047}{{\em JHEP}
  {\bfseries 06} (2010) 047},
\href{http://arxiv.org/abs/1001.5443}{{\ttfamily arXiv:1001.5443 [hep-ph]}}.

\bibitem{Dermisek:2006qj}
R.~Dermisek, H.~D. Kim, and I.-W. Kim, ``{Mediation of supersymmetry breaking
  in gauge messenger models},'' {\em JHEP} {\bfseries 10} (2006) 001,
\href{http://arxiv.org/abs/hep-ph/0607169}{{\ttfamily arXiv:hep-ph/0607169}}.

\bibitem{Giudice:1997ni}
G.~F. Giudice and R.~Rattazzi, ``{Extracting Supersymmetry-Breaking Effects
  from Wave- Function Renormalization},''
  \href{http://dx.doi.org/10.1016/S0550-3213(97)00647-0}{{\em Nucl. Phys.}
  {\bfseries B511} (1998) 25--44},
\href{http://arxiv.org/abs/hep-ph/9706540}{{\ttfamily arXiv:hep-ph/9706540}}.

\bibitem{McGarrie:2010kh}
M.~McGarrie and R.~Russo, ``{General Gauge Mediation in 5D},''
\href{http://arxiv.org/abs/1004.3305}{{\ttfamily arXiv:1004.3305 [hep-ph]}}.

\bibitem{Seiberg:2008qj}
N.~Seiberg, T.~Volansky, and B.~Wecht, ``{Semi-direct Gauge Mediation},''
  \href{http://dx.doi.org/10.1088/1126-6708/2008/11/004}{{\em JHEP} {\bfseries
  11} (2008) 004},
\href{http://arxiv.org/abs/0809.4437}{{\ttfamily arXiv:0809.4437 [hep-ph]}}.

\bibitem{Benakli:2010gi}
K.~Benakli and M.~D. Goodsell, ``{Dirac Gauginos, Gauge Mediation and
  Unification},''
\href{http://arxiv.org/abs/1003.4957}{{\ttfamily arXiv:1003.4957 [hep-ph]}}.

\bibitem{Nelson:1993nf}
A.~E. Nelson and N.~Seiberg, ``{R symmetry breaking versus supersymmetry
  breaking},'' \href{http://dx.doi.org/10.1016/0550-3213(94)90577-0}{{\em Nucl.
  Phys.} {\bfseries B416} (1994) 46--62},
\href{http://arxiv.org/abs/hep-ph/9309299}{{\ttfamily arXiv:hep-ph/9309299}}.

\bibitem{Shih:2007av}
D.~Shih, ``{Spontaneous R-symmetry breaking in O'Raifeartaigh models},''
  \href{http://dx.doi.org/10.1088/1126-6708/2008/02/091}{{\em JHEP} {\bfseries
  02} (2008) 091},
\href{http://arxiv.org/abs/hep-th/0703196}{{\ttfamily arXiv:hep-th/0703196}}.

\bibitem{Komargodski:2009jf}
Z.~Komargodski and D.~Shih, ``{Notes on SUSY and R-Symmetry Breaking in
  Wess-Zumino Models},''
  \href{http://dx.doi.org/10.1088/1126-6708/2009/04/093}{{\em JHEP} {\bfseries
  04} (2009) 093},
\href{http://arxiv.org/abs/0902.0030}{{\ttfamily arXiv:0902.0030 [hep-th]}}.

\bibitem{Raby:2006sk}
S.~Raby, ``{Grand Unified Theories},''
\href{http://arxiv.org/abs/hep-ph/0608183}{{\ttfamily arXiv:hep-ph/0608183}}.

\bibitem{Raby:2008gh}
S.~Raby, ``{SUSY GUT Model Building},''
  \href{http://dx.doi.org/10.1140/epjc/s10052-008-0736-x}{{\em Eur. Phys. J.}
  {\bfseries C59} (2009) 223--247},
\href{http://arxiv.org/abs/0807.4921}{{\ttfamily arXiv:0807.4921 [hep-ph]}}.

\bibitem{Abel:2009bj}
S.~Abel and V.~V. Khoze, ``{Dual unified SU(5)},''
  \href{http://dx.doi.org/10.1007/JHEP01(2010)006}{{\em JHEP} {\bfseries 01}
  (2010) 006},
\href{http://arxiv.org/abs/0909.4105}{{\ttfamily arXiv:0909.4105 [hep-ph]}}.

\bibitem{Abel:2008tx}
S.~Abel and V.~V. Khoze, ``{Direct Mediation, Duality and Unification},''
  \href{http://dx.doi.org/10.1088/1126-6708/2008/11/024}{{\em JHEP} {\bfseries
  11} (2008) 024},
\href{http://arxiv.org/abs/0809.5262}{{\ttfamily arXiv:0809.5262 [hep-ph]}}.

\bibitem{ArkaniHamed:1998kj}
N.~Arkani-Hamed, G.~F. Giudice, M.~A. Luty, and R.~Rattazzi,
  ``{Supersymmetry-breaking loops from analytic continuation into
  superspace},'' \href{http://dx.doi.org/10.1103/PhysRevD.58.115005}{{\em Phys.
  Rev.} {\bfseries D58} (1998) 115005},
\href{http://arxiv.org/abs/hep-ph/9803290}{{\ttfamily arXiv:hep-ph/9803290}}.

\bibitem{Ray:2006wk}
S.~Ray, ``{Some properties of meta-stable supersymmetry-breaking vacua in
  Wess-Zumino models},''
  \href{http://dx.doi.org/10.1016/j.physletb.2006.09.009}{{\em Phys. Lett.}
  {\bfseries B642} (2006) 137--141},
\href{http://arxiv.org/abs/hep-th/0607172}{{\ttfamily arXiv:hep-th/0607172}}.

\bibitem{Murayama:2006yf}
H.~Murayama and Y.~Nomura, ``{Gauge mediation simplified},''
  \href{http://dx.doi.org/10.1103/PhysRevLett.98.151803}{{\em Phys. Rev. Lett.}
  {\bfseries 98} (2007) 151803},
\href{http://arxiv.org/abs/hep-ph/0612186}{{\ttfamily arXiv:hep-ph/0612186}}.

\bibitem{Murayama:2007fe}
H.~Murayama and Y.~Nomura, ``{Simple scheme for gauge mediation},''
  \href{http://dx.doi.org/10.1103/PhysRevD.75.095011}{{\em Phys. Rev.}
  {\bfseries D75} (2007) 095011},
\href{http://arxiv.org/abs/hep-ph/0701231}{{\ttfamily arXiv:hep-ph/0701231}}.

\bibitem{Kitano:2006xg}
R.~Kitano, H.~Ooguri, and Y.~Ookouchi, ``{Direct mediation of meta-stable
  supersymmetry breaking},''
  \href{http://dx.doi.org/10.1103/PhysRevD.75.045022}{{\em Phys. Rev.}
  {\bfseries D75} (2007) 045022},
\href{http://arxiv.org/abs/hep-ph/0612139}{{\ttfamily arXiv:hep-ph/0612139}}.

\bibitem{Witten:1981kv}
E.~Witten, ``{Mass Hierarchies in Supersymmetric Theories},''
\href{http://dx.doi.org/10.1016/0370-2693(81)90885-6}{{\em Phys. Lett.}
  {\bfseries B105} (1981) 267}.

\bibitem{Matos:2009xv}
L.~F. Matos, ``{Some examples of F and D-term SUSY breaking models},''
\href{http://arxiv.org/abs/0910.0451}{{\ttfamily arXiv:0910.0451 [hep-ph]}}.

\bibitem{Abel:2006my}
S.~A. Abel, J.~Jaeckel, and V.~V. Khoze, ``{Why the early universe preferred
  the non-supersymmetric vacuum. II},'' {\em JHEP} {\bfseries 01} (2007) 015,
\href{http://arxiv.org/abs/hep-th/0611130}{{\ttfamily arXiv:hep-th/0611130}}.

\bibitem{Abel:2006cr}
S.~A. Abel, C.-S. Chu, J.~Jaeckel, and V.~V. Khoze, ``{SUSY breaking by a
  metastable ground state: Why the early universe preferred the
  non-supersymmetric vacuum},'' {\em JHEP} {\bfseries 01} (2007) 089,
\href{http://arxiv.org/abs/hep-th/0610334}{{\ttfamily arXiv:hep-th/0610334}}.

\bibitem{Craig:2006kx}
N.~J. Craig, P.~J. Fox, and J.~G. Wacker, ``{Reheating metastable
  O'Raifeartaigh models},''
  \href{http://dx.doi.org/10.1103/PhysRevD.75.085006}{{\em Phys. Rev.}
  {\bfseries D75} (2007) 085006},
\href{http://arxiv.org/abs/hep-th/0611006}{{\ttfamily arXiv:hep-th/0611006}}.

\bibitem{Fischler:2006xh}
W.~Fischler, V.~Kaplunovsky, C.~Krishnan, L.~Mannelli, and M.~A.~C. Torres,
  ``{Meta-Stable Supersymmetry Breaking in a Cooling Universe},'' {\em JHEP}
  {\bfseries 03} (2007) 107,
\href{http://arxiv.org/abs/hep-th/0611018}{{\ttfamily arXiv:hep-th/0611018}}.

\bibitem{Dimopoulos:1997ww}
S.~Dimopoulos, G.~R. Dvali, R.~Rattazzi, and G.~F. Giudice, ``{Dynamical soft
  terms with unbroken supersymmetry},''
  \href{http://dx.doi.org/10.1016/S0550-3213(97)00603-2}{{\em Nucl. Phys.}
  {\bfseries B510} (1998) 12--38},
\href{http://arxiv.org/abs/hep-ph/9705307}{{\ttfamily arXiv:hep-ph/9705307}}.

\bibitem{Murayama:1997pb}
H.~Murayama, ``{A Model of direct gauge mediation},''
  \href{http://dx.doi.org/10.1103/PhysRevLett.79.18}{{\em Phys. Rev. Lett.}
  {\bfseries 79} (1997) 18--21},
\href{http://arxiv.org/abs/hep-ph/9705271}{{\ttfamily arXiv:hep-ph/9705271}}.

\bibitem{Nardecchia:2009ew}
M.~Nardecchia, A.~Romanino, and R.~Ziegler, ``{Tree Level Gauge Mediation},''
  \href{http://dx.doi.org/10.1088/1126-6708/2009/11/112}{{\em JHEP} {\bfseries
  11} (2009) 112},
\href{http://arxiv.org/abs/0909.3058}{{\ttfamily arXiv:0909.3058 [hep-ph]}}.

\bibitem{FileviezPerez:2009im}
P.~Fileviez~Perez, H.~Iminniyaz, G.~Rodrigo, and S.~Spinner, ``{Gauge Mediated
  SUSY Breaking via Seesaw},''
  \href{http://dx.doi.org/10.1103/PhysRevD.81.095013}{{\em Phys. Rev.}
  {\bfseries D81} (2010) 095013},
\href{http://arxiv.org/abs/0911.1360}{{\ttfamily arXiv:0911.1360 [hep-ph]}}.

\bibitem{Dimopoulos:1981dw}
S.~Dimopoulos, S.~Raby, and F.~Wilczek, ``{Proton Decay in Supersymmetric
  Models},''
\href{http://dx.doi.org/10.1016/0370-2693(82)90313-6}{{\em Phys. Lett.}
  {\bfseries B112} (1982) 133}.

\bibitem{Ibanez:1982fr}
L.~E. Ibanez and G.~G. Ross, ``{SU(2)-L x U(1) Symmetry Breaking as a Radiative
  Effect of Supersymmetry Breaking in Guts},''
\href{http://dx.doi.org/10.1016/0370-2693(82)91239-4}{{\em Phys. Lett.}
  {\bfseries B110} (1982) 215--220}.

\bibitem{Banks:1982mg}
T.~Banks and V.~Kaplunovsky, ``{NOSONOMY OF AN UPSIDE DOWN HIERARCHY MODEL.
  1},''
\href{http://dx.doi.org/10.1016/0550-3213(83)90113-X}{{\em Nucl. Phys.}
  {\bfseries B211} (1983) 529}.

\end{thebibliography}\endgroup
\end{document}